\documentclass[aps,prl,showpacs,twocolumn,superscriptaddress,floatfix]{revtex4}
\pdfoutput=1
\usepackage{graphicx}
\usepackage{amssymb,amsmath} 
\usepackage{color}
\usepackage{multirow}
\usepackage[normalem]{ulem}
\usepackage{nicefrac}
\usepackage{pifont}

\RequirePackage[xcolornames,svgnames,dvipsnames,rgb]{xcolor}
\pdfpageattr {/Group << /S /Transparency /I true /CS /DeviceRGB>>}

\newcommand\T{\rule{0pt}{2.6ex}}
\newcommand\B{\rule[-1.2ex]{0pt}{0pt}}
\newcommand{\dxz}{$d_{xz}$}
\newcommand{\dyz}{$d_{yz}$}
\newcommand{\dxy}{$d_{xy}$}
\newcommand{\pz}{$p_{z}$}
\newcommand{\dz}{$d_{z^2}$}

\newcommand{\Ctopone}{C$_{\mathrm{top-1}}$}
\newcommand{\Ctoptwo}{C$_{\mathrm{top-2}}$}
\newcommand{\Cbridge}{C$_{\mathrm{bridge}}$}

\begin{document}

\title{Atomic-scale inversion of spin polarization at an organic-antiferromagnetic interface}

\author{Nuala M.~Caffrey}
\email[Electronic address: ]{caffrey@theo-physik.uni-kiel.de}
\affiliation{Institut f\"ur Theoretische Physik und Astrophysik, Christian-Albrecht-Universit\"at zu Kiel, D-24098 Kiel, Germany}
\author{Paolo Ferriani}
\affiliation{Institut f\"ur Theoretische Physik und Astrophysik, Christian-Albrecht-Universit\"at zu Kiel, D-24098 Kiel, Germany}
\author{Simone Marocchi}
\affiliation{Dipartimento di Fisica, Universit\'a di Modena e Reggio Emilia, Via Campi 213/A, 41125 Modena, Italy.}
\affiliation{S3 - Istituto di Nanoscienze - CNR, Via Campi 213/A, 41125 Modena, Italy}
\author{Stefan Heinze}
\affiliation{Institut f\"ur Theoretische Physik und Astrophysik, Christian-Albrecht-Universit\"at zu Kiel, D-24098 Kiel, Germany}

\date{\today}

\begin{abstract}
Using first-principles calculations, we show that the magnetic properties of a two-dimensional antiferromagnetic
transition-metal surface are modified on the atomic scale by the adsorption of small organic molecules.
We consider benzene (C$_6$H$_6$), cyclooctatetraene (C$_8$H$_8$) and a small transition metal - benzene complex (BzV) adsorbed
on a single atomic layer of Mn deposited on the W(110) surface -- a surface which exhibits a nearly antiferromagnetic alignment of
the magnetic moments in adjacent Mn rows.
Due to the spin-dependent hybridization of the molecular $p_z$ orbitals with the $d$ states of the Mn monolayer
there is a significant reduction of the magnetic moments in the Mn film.
Furthermore, the spin-polarization at this organic-antiferromagnetic interface is found to be modulated on the atomic scale, both enhanced and
inverted, as a result of the molecular adsorption. 
We show that this effect can be resolved by spin-polarized scanning tunneling microscopy (SP-STM).
Our simulated SP-STM images display a spatially-dependent spin-resolved vacuum charge density above an adsorbed molecule -- i.e., different regions above the molecule sustain different signs of spin polarization.
While states with $s$ and $p$ symmetry dominate the vacuum charge density in the vicinity of the Fermi energy for the clean magnetic surface, we demonstrate that after
a molecule is adsorbed those $d$-states, which are normally suppressed due to their symmetry, can play a crucial role in the vacuum due to
their interaction with the molecular orbitals.
We also model the effect of small deviations from perfect antiferromagnetic ordering, induced by the slight canting of magnetic moments due to the spin spiral ground state of Mn/W(110).

\end{abstract}


\maketitle

\section{Introduction}
The emerging field of organic spintronics aims to combine the advantages of molecular electronics, such as device miniaturisation and fabrication ease, with the massive potential for application inherent in spintronics \cite{BoganiReview2008, Dediu2009, SanvitoReview2011}. Organic materials are particularly promising in spintronic devices as they exhibit weak spin-orbit and hyperfine interactions, resulting in long spin coherence times \cite{Dediu2002, Pramanik2007}. Moreover, they have a high chemical diversity and offer the possibility for selectively tuning their electronic and magnetic properties via ligand modification. The first reported organic spin valve used a thin layer of Alq$_3$ as an organic spacer between LSMO and Co electrodes \cite{Xiong2004}. Recently, a magnetoresistance of 300\% was found for the same structure \cite{Barraud2010}.
In an effort to meet the demand for miniaturisation there has been a drive to use single organic molecules as spacer elements \cite{Rocha2005, Haiss2006, Tao2006, Bogani2008}. An added advantage is that the resistance of such a device would be considerably smaller than that of a thin-film based device. Very high magnetoresistance effects have already been predicted \cite{Sanvito2006, PhysRevLett.96.166804, PhysRevB.76.024438, Rocha2007, Saffarzadeh2008} and achieved \cite{Schmaus2011, Urdampilleta2011} in such single molecule spin-valves for both the tunnelingand conducting regime.

In many cases it is the interface between the organic molecule and the inorganic electrodes that determines the spin injection properties. The exchange-split densities of states of the electrodes result in the spin-selective shift and broadening of the molecular orbitals. The result is a spin-polarized interface, the so-called `spinterface' \cite{Sanvito2010}. The spin-split molecular orbitals \cite{Kawahara2012, Schwoebel2012} as well as the spin-polarized interface itself \cite{Djeghloul2013} have been directly observed.
Even a single organic molecule has been demonstrated to be capable of locally manipulating the spin polarization emerging from a clean magnetic surface.
This was shown by Atodiresei et al. \cite{PhysRevLett.105.066601} and Brede et al. \cite{PhysRevLett.105.047204} who considered the effect the adsorption of small organic molecules can have on a ferromagnetic Fe/W(110) surface. They found that even a molecule as small as a benzene ring (Bz) can act as an efficient spin filter, capable of inverting the spin polarization of the surface, due to the particular hybridization of orbitals at the interface -- an effect which has also been observed for single atoms adsorbed on magnetic surfaces \cite{PhysRevB.82.012409,Ferriani2010}.

Most organic-inorganic interfaces studied to date involve ferromagnetic surfaces \cite{Wende2007, Javaid2010, PhysRevB.82.094443, Lach2012, Raman2013, PhysRevB.87.054420}. More recently, the
adsorption and coupling of molecules on substrates with layered antiferromagnetic order has been used
to explore spin-polarized transport and the exchange bias effect \cite{Rizzini2012,Bagrets2012}.
In this work, we consider a surface with a more complex spin structure -- a monolayer of Mn on the W(110) surface -- which
has been characterized on the atomic scale \cite{PhysRevB.66.014425,Bode2007}. It has been found to
exhibit a spin spiral ground state propagating along the [$1\bar{1}0$] direction with an angle of $\sim$173$^\circ$ between magnetic moments on adjacent rows \cite{Bode2007}. This means that neighbouring rows of Mn atoms have magnetic moments approximately opposite in direction but with an orientation relative to the film that varies slowly across the surface. Locally one can therefore approximate it
as a two-dimensional antiferromagnet. This surface has already been used as as a magnetic template in order to determine and manipulate the
spin direction of individual Co adatoms on the atomic scale \cite{Serrate2010}. Here, we explore the effect of single molecules on its local magnetic properties.

We present a density functional theory study of the structural, electronic and magnetic properties of several simple organic molecules adsorbed on the Mn/W(110) surface. Due to the long period of the spin spiral, we consider it as approximately collinear to model its interaction with localized adsorbed molecules. 
The molecules under investigation include benzene  (Bz: C$_6$H$_6$), cyclooctatetraene (COT: C$_8$H$_8$) and the small benzene - transition metal complex BzV. 
We show that due to hybridization with the molecule the magnetic moments of the adjacent Mn atoms can be strongly reduced
and that the spin polarization of such an organic-antiferromagnetic interface is modulated on the atomic scale and can be
enhanced and inverted with respect to the clean Mn film.

The modification of the spin polarization due to this interface can be revealed by spin-polarized scanning tunneling microscopy (SP-STM) as demonstrated by our calculations.
We find that the exact magnitude and sign of the spin polarization in the vacuum above the molecule is strongly dependent on the bonding details at the interface and, due to the antiferromagnetic surface, exhibits a strong intra-molecular spatial dependence. In almost all cases, the sign of the vacuum spin polarization above the adsorbed molecule is inverted and its magnitude enhanced. We demonstrate that this modification of the charge density in the vacuum is a hybridization-induced effect between the $p$-orbitals of the organic molecule and the spin-split $d$-orbitals of the magnetic surface.
The effect of the non-collinear spin structure of the Mn monolayer on the spin-polarized STM images is also calculated, using a model based on that of Tersoff and Hamann \cite{PhysRevLett.86.4132}.

The paper is organized as follows. After briefly describing the computational method, the structural details of the adsorbed molecules are discussed. This is followed by a description of the electronic and magnetic properties of the hybrid organic - inorganic interface. In particular, we discuss the spin resolved local density of states as calculated in the vacuum above the molecules, i.e. the simulated STM images. Finally, we consider in an approximate way the effect the true spin spiral ground state can have on the simulated SP-STM images. The final section summarizes our main conclusions.

\section{Methods}
In this work, density functional theory calculations are performed using the {\sc vasp} code \cite{Kresse1996, Kresse1999}. The Perdew-Burke-Ernzerhof (PBE) \cite{Perdew1996} parametrization of the generalized gradient approximation (GGA) is employed. The projector-augmented wave (PAW) method \cite{PhysRevB.50.17953} is used with the standard PAW potentials supplied with the VASP distribution. The plane wave basis set was converged using a 400~eV energy cutoff. A 6 $\times$ 6 $\times$ 1 k-point Monkhorst-Pack mesh \cite{PhysRevB.13.5188} was used to sample the three-dimensional Brillouin zone. A Gaussian smearing of 0.01~eV was used for the initial occupations.
Constant-height spin polarized scanning tunneling microscopy (SP-STM) images were simulated using the the method of Tersoff and Hamann \cite{PhysRevB.31.805} and its generalization to SP-STM \cite{PhysRevLett.86.4132}.
The central quantity of this scheme is the spin dependent local density of states in the vacuum integrated over a particular energy range corresponding to an experimental bias voltage.

The substrate is modelled using a 3 layer slab of W atoms and one layer of Mn atoms in the collinear checkerboard antiferromagnetic configuration. The GGA calculated lattice constant of W is 3.17 \AA, in good agreement with the experimental value of 3.165 \AA. The size of the supercell was chosen such that a distance of at least 12.5 \AA\ was maintained between the centre of neighbouring molecules. Additionally, a thick vacuum layer of approximately 3~nm was included in the direction normal to the surface to ensure no spurious interactions between repeating slabs.
The relaxed interlayer distance between Mn and W is 2.08 \AA\ whereas the interlayer distance between two layers of W is 2.24 \AA. The energy of the $c$(2$\times$2) AFM configuration is 187~meV/atom and 109~meV/atom lower than the ferromagnetic and the $p$(2$\times$1) AFM configurations, respectively. These energy differences were calculated for a symmetric slab consisting of seven layers of W(110) and a Mn layer on both sides and compare favourable to previous results in the literature \cite{PhysRevB.66.014425}.

\section{Structures}
\begin{figure*}[ht]
\begin{centering}
\includegraphics[width=0.95\linewidth]{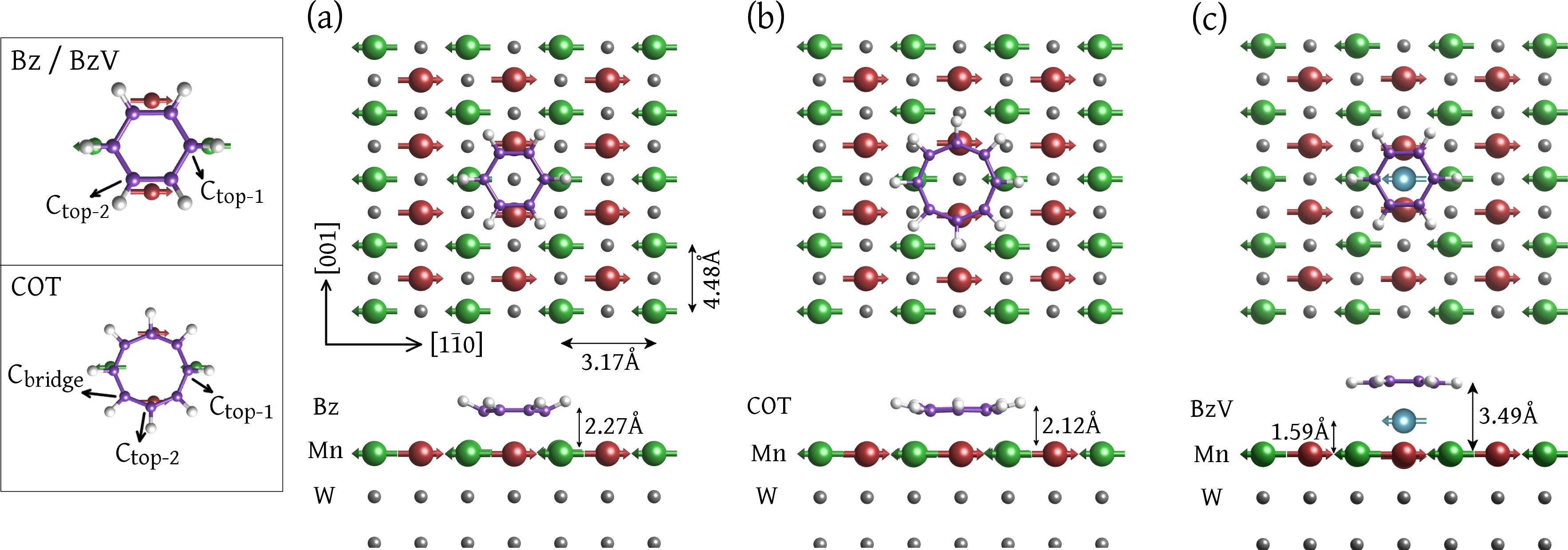}
\caption{\label{fig:structures}(Color online) Top and side view of the adsorption geometry of a (a) Bz molecule, (b) COT molecule and (c) BzV molecule adsorbed on a Mn/W(110) surface. A c(6 $\times$ 8) surface unit cell was employed for the case of Bz and BzV adsorption while a larger surface unit cell of c(8 $\times$ 10) was used for the adsorption of the COT molecule (only a portion of this cell is shown). The antiferromagnetic alignment of Mn spins is depicted using green and red spheres and arrows.}
\end{centering}
\end{figure*}
Several initial molecular adsorption configurations are considered for all molecules, including hollow, bridge and top sites. The most energetically stable molecular position is, in all cases, found to be over the hollow site of the Mn layer (Fig.~\ref{fig:structures}). Rotations of the adsorbed molecule were also considered. Fig.~\ref{fig:structures} shows the most energetically favourable configurations. The ionic positions of the molecular atoms and the Mn layer were optimized until all residual forces were less than 0.01~eV/\AA. In all cases the coordinates of the W atoms were held fixed.
The effect of including dispersion forces was also considered using the semi-empirical approach of Grimme \cite{Grimme2006} as implemented in the {\sc vasp} code \cite{Bucko2010}. For the case of the benzene molecule, the inclusion of dispersion forces were found to decrease the binding distance of the molecule to the surface by 0.04~\AA. However, this was not found to have any appreciable effect on the 
spin polarization at the interface or the simulated STM images and so was neglected in all further calculations.

Fig.~\ref{fig:structures}(a) shows a top and side view of the optimized Bz/Mn/W(110) structure. After relaxation the molecular plane is no longer flat; the hydrogen atoms lie further from the surface than the carbon atoms by a distance of 0.41~\AA. The shortest C -- Mn bond length is 2.10~\AA\ which occurs when the carbon atom sits directly on top of a Mn ion. These carbon atoms are referred to as \Ctopone (see Fig.~\ref{fig:structures}).
Additionally, the C -- C bonds in Bz are no longer equal and are larger compared to the isolated molecule (1.456~\AA\ and 1.419~\AA\ compared to 1.398~\AA), in order to facilitate its adsorption onto the surface lattice. The carbon - hydrogen bonds remain at their isolated molecule length of 1.09~\AA.

In contrast to planar Bz, the isolated COT molecule exists in a ``tub-like'' conformation \cite{Kaufman1948} in order to reduce angle strain. However, when adsorbed on the surface considered here, the planar structure of the carbon atoms is approximately restored. This has previously been shown to occur when COT is chemisorbed onto a surface with the resulting strong hybridization between molecular and surface electronic states \cite{Harutyunyan2013}. Similar to the Bz case, the hydrogen atoms are positioned in a plane higher than that of the carbons, in this case by approximately 0.11~\AA.
The C -- C bond lengths, while alternating in the isolated molecule between 1.47~\AA\ and 1.35~\AA, become approximately equal when adsorbed on the surface, at 1.43~\AA. As with Bz, the COT molecule prefers to bind at a hollow site of the Mn lattice. Due to the symmetry of the adsorption the C atoms can be grouped into three different types; two carbons that sit atop Mn atoms with negative spin polarization, two carbons that sit atop Mn atoms with positive spin polarization, and four carbons that lie across a Mn --Mn bridge site (see Fig.~\ref{fig:structures}). These are denoted \Ctopone, \Ctoptwo\ and \Cbridge, respectively. Note that due to the slightly asymmetric binding of the molecule to the surface the division into such categories is only approximate.

The final molecule studied introduces a magnetic adatom, in this case Vanadium, between the molecule and the surface. The resultant molecule is then a half-sandwich, representing an ideal candidate for the smallest possible organometallic molecular magnet. The interaction of 3$d$ transition metal ions with Bz molecules have been previously studied both experimentally \cite{Jaeger2004, Duncan2008} and theoretically \cite{kandalam:10414,  Mokrousov2007}.
The distance between the benzene ring and the surface has now increased to 3.49~\AA\ due to the V atom which is positioned over a hollow site in the Mn lattice and 1.59~\AA\ above it.

\section{Results and Discussion}

\subsection{Electronic structure}
\begin{figure}
\begin{centering}
\includegraphics[width=0.95\linewidth]{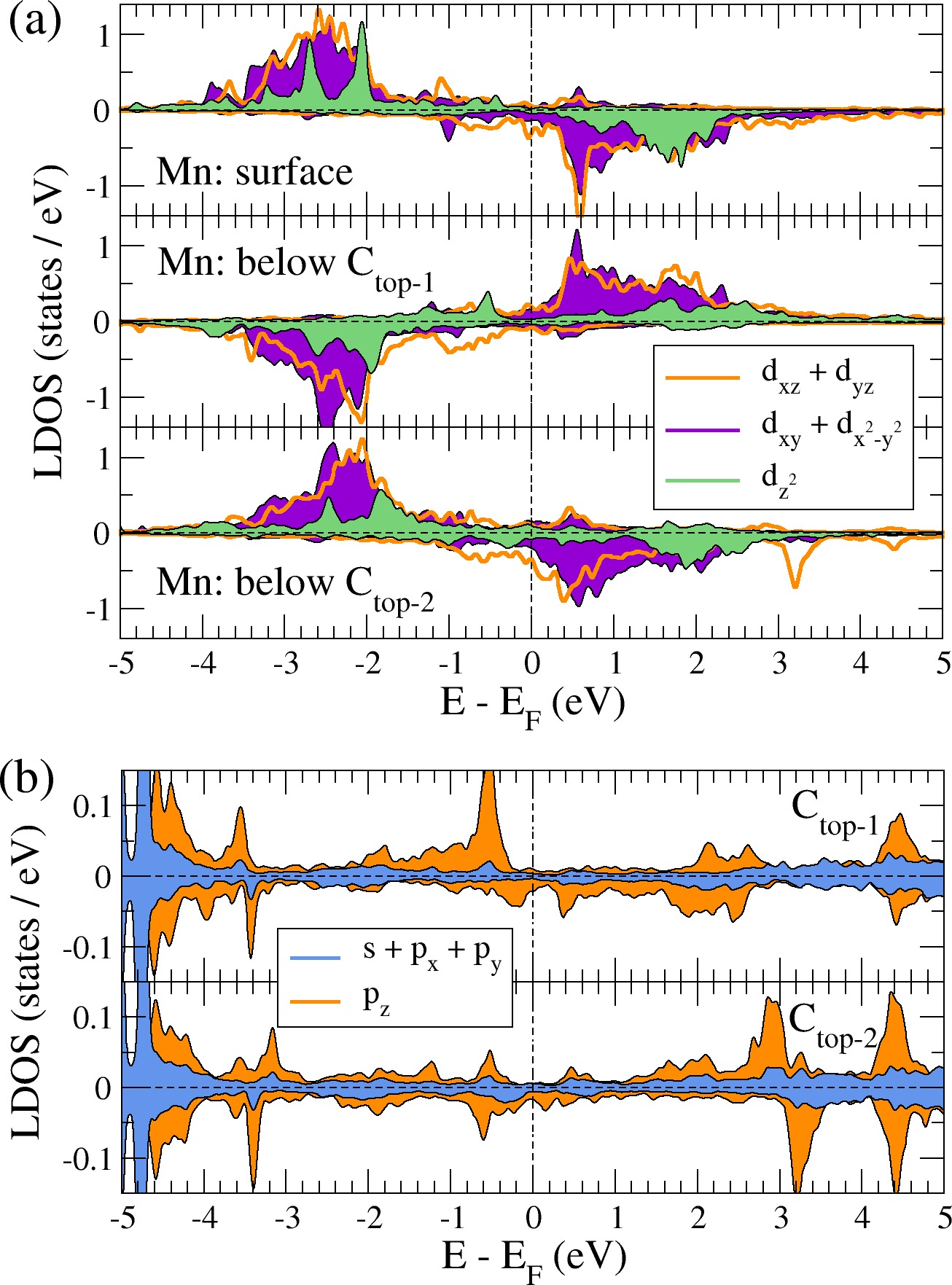}
\caption{\label{fig:bz_dos}(Color online) The spin resolved local density of states for a Bz molecule adsorbed onto a Mn/W(110) surface, showing (a) states from the Mn layer and (b) states from the C atoms of the Bz molecule. The positive and negative values of the vertical axis correspond to spin-up and spin-down, respectively. Refer to Fig.~\ref{fig:structures}(a) for the definition of the atom types.
}
\end{centering}
\end{figure}

\begin{figure}
\begin{centering}
\includegraphics[width=0.95\linewidth]{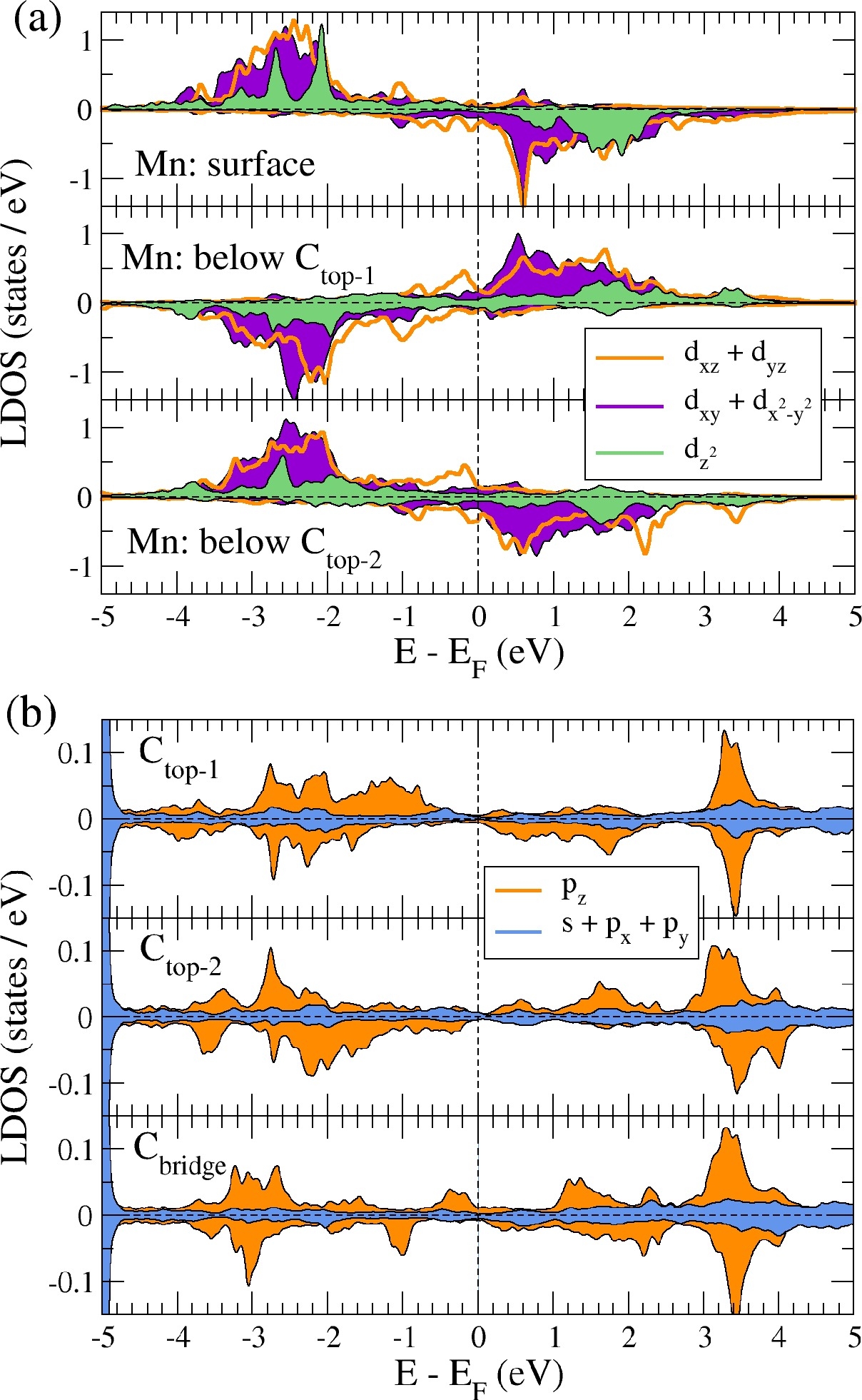}
\caption{\label{fig:cot_dos}(Color online) The spin resolved local density of states for a COT molecule adsorbed onto a Mn/W(110) surface, showing (a) the Mn layer and (b) the C atoms of the COT molecule.}
\end{centering}
\end{figure}

\begin{figure}
\begin{centering}
\includegraphics[width=0.95\linewidth]{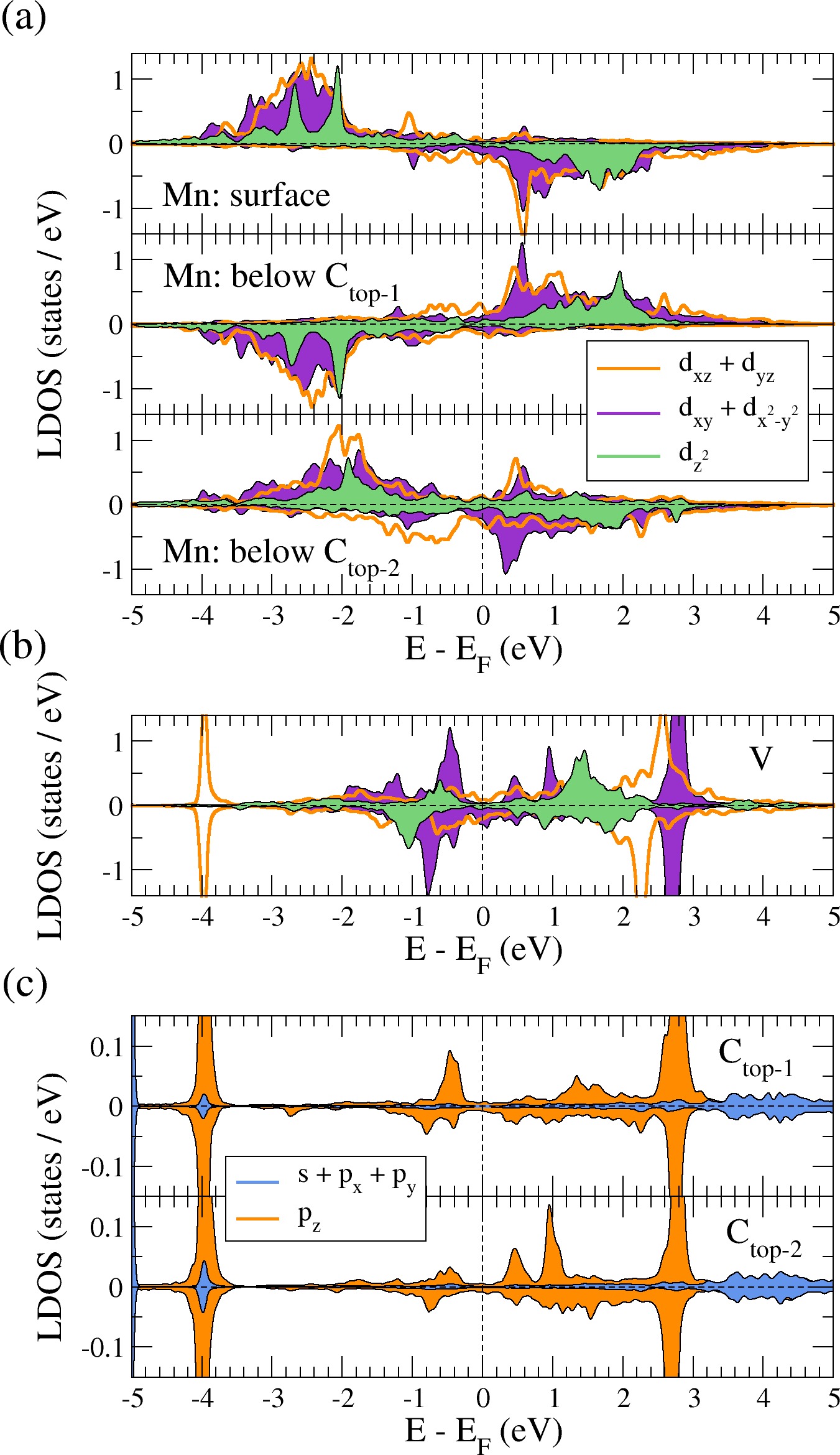}
\caption{\label{fig:bzV_dos}(Color online) The spin resolved local density of states for a BzV molecule adsorbed onto a Mn/W(110) surface, showing (a) the Mn layer, (b) the V atom and (c) the C atoms of the BzV molecule.}
\end{centering}
\end{figure}

In order to illustrate the binding mechanism and hybridization between the organic molecules and the inorganic antiferromagnetic surface, the local density of states (LDOS) are shown in Fig.~\ref{fig:bz_dos} (Bz),  Fig.~\ref{fig:cot_dos} (COT) and Fig.~\ref{fig:bzV_dos} (BzV).
The upper panel of Fig.~\ref{fig:bz_dos}(a) shows the LDOS of a Mn atom far from the influence of the molecule with a magnetic moment
pointing to the left (green atom in Fig.~\ref{fig:structures}) which in our definition of the global spin quantization axis (SQA) exhibits
a greater spin down density of states close to the Fermi level.
In an energy window relevant to STM measurements, the dominant states are spin down with \dxz\ symmetry.
The density of spin up states in this energy range is much lower and has contributions from all $d$ orbitals as well as $s$ states.
Comparing the LDOS of this Mn atom to the LDOS of the Mn atoms located directly below Bz it becomes evident that states with a component in the direction normal to the plane of the surface, i.e., \dz, \dxz\ and \dyz, are most altered by the adsorption of the molecule due to the formation of hybrid $p$ -- $d$ bonds. States with in-plane components are not altered significantly.
Fig.~\ref{fig:bz_dos}(b) shows the DOS of \Ctopone\ and \Ctoptwo. For the case of \Ctopone\ one can see a strong hybridization between the Mn \dz\ states and the C \pz\ states at $-0.5$~eV.
For the case of the Mn atom beneath a C -- C bond, the \dxz\ and \dyz\ states play a marked role.
This can be seen at $-0.6$~eV and  $+0.2$~eV for the spin down \pz\ states of \Ctoptwo\ and the \dxz\ states of the Mn below \Ctoptwo. A similarly large overlap can be seen at  $+3.2$~eV.
The hybridization between the non-magnetic molecule and the magnetic surface has important consequences on the spin polarization of the molecule. Examining the LDOS close to the Fermi level, one can see that the spin polarization of the carbon atoms is not only significant but that it varies in magnitude and sign from carbon atom to carbon atom. Consider, for instance, the LDOS of \Ctopone\ and \Ctoptwo\ at  $-0.6$~eV. Here, the spin polarization of \Ctopone\ is positive, while that of \Ctoptwo\ is both much smaller in magnitude and of opposite sign.

For the case of the COT molecule (Fig.~\ref{fig:cot_dos}), there is a Mn atom, albeit with opposite directions of the magnetic moment, directly below both \Ctopone\ and \Ctoptwo. This is reflected in the very similar DOS for both types of carbon atoms. The remaining four carbons are positioned above a Mn -- Mn bond (\Cbridge). As for the case of the Bz molecule, the \dxz\ and \dyz\ orbitals of Mn atoms below these carbon atoms contribute most to the hybridization. Of particular importance are the spin up \dxz\ states which display a broad peak just below the Fermi level at $-0.2$~eV. Due to hybridization, this broad peak can also be seen in the spin up \Cbridge\ \pz\ states. This state dominates at the Fermi level and leads to a positive spin polarization, an important fact when considering the STM images later.

Finally, the LDOS for the case of the BzV molecule adsorbed on the Mn/W(110) surface is shown in Fig.~\ref{fig:bzV_dos}.
The spin dependent hybridization now occurs between the Mn layer and the molecular orbitals of the V-Bz complex. In 
the isolated BzV molecule, the dominant states close to the Fermi level comprise of the doubly degenerate $\delta$ states \cite{Mokrousov2007}. Upon adsorption on the magnetic surface, the spin down $\delta$ states hybridize with the spin down \dxz\ orbitals of the Mn atoms below \Ctoptwo. Furthermore, the unoccupied spin up $\delta$ states are hybridized with the \dxz\ states of the Mn atom below \Ctopone.
Looking now at the DOS of \Ctopone\ and \Ctoptwo, it is clear that close to the Fermi level the carbon \pz\ orbitals are strongly hybridized with the spin split in-plane vanadium orbitals
of \dxy\ and $d_{x^2-y^2}$ symmetry.
Below $-4$~eV, the $\pi$, L$s$ and s$\sigma$ molecular orbitals, which are located mostly on the carbon sites of the Bz ring, are not involved in the hybridization with the Mn layer and so remain unchanged.

\subsection{Interfacial effects and magnetism}
The hybridization between states of the molecule and the surface results in a strong modification of the interfacial properties of the system and in particular the magnetic properties of both the surface and the adsorbents.
\begin{table}
\begin{centering}
\begin{tabular}{lccccc}%
\multicolumn{6}{c}{}\\
\hline \hline
							\T \B 		&  	& Bz 		&  	& COT 		& 	BzV		\\
\cline{1-6}
{\small Mn (clean surface)} 	\T \B		&	&	$+3.37$	&	&	$+3.38$	&	$3.38$	\\
{\small Mn (below \Ctopone)}	\T \B		&	&	$-3.09$	&	&	$-2.94$	&	$-3.08$	\\
{\small Mn (below \Ctoptwo)}	\T \B		&	&	$+2.66$	&	&	$+2.83$	&	$+1.57$	\\
{\small \Ctopone}			\T \B		&	&	$+0.08$	&	&	$+0.05$	&	$+0.01$	\\
{\small \Ctoptwo}			\T \B		&	&	$-0.03$	&	&	$-0.04$	&	$-0.01$	\\
{\small \Cbridge}				\T \B		&	&			&	&	$+0.01$	&			\\
{\small V}					\T \B		&	&			&	&			&	$-0.31$	\\
\hline
\end{tabular}
\caption{\label{tab:magnetic_moments}\,Magnetic moments of the Mn atoms and the C atoms in $\mu_{\rm B}$ of the Bz, COT and BzV molecules adsorbed on the Mn/W(110) surface.}
\end{centering}
\end{table}
The magnetic moments of the Mn atoms and the induced moments on the organic molecules are shown in Table~\ref{tab:magnetic_moments}. The magnetic moment of a Mn atom on a clean Mn/W(110) surface is $\pm3.4$~$\mu_B$. This moment is modified by the adsorption of an organic molecule due to the hybridization of the $d$ states with the $\pi$ orbitals. In the case of all three molecules studied here the Mn magnetic moments are significantly reduced, with the reduction dependent on the details of the bonding. The largest drop in moment occurs for the Mn atom beneath the \Ctoptwo\ atoms, by  $0.71$~$\mu_B$, $0.55$~$\mu_B$ and $1.81$~$\mu_B$ for Bz, COT and BzV, respectively.
The largest reduction in surface magnetic moment occurs for Mn atoms with two neighbouring C atoms. If only one C atom is bound to the Mn atom, then the largest decrease in moment occurs when the bond between the Mn atom and the C atom is shortest. For instance, the reduction in moment below \Ctopone\ and \Ctoptwo\ for the case of the COT molecule is smaller than that below  \Ctoptwo\ for the case of the Bz molecule due to the larger Mn-C bond length for the adsorbed COT molecule. We highlight in particular the large reduction of the surface magnetic moment for the case of the magnetic molecule, in particular for the Mn atom bound to two C atoms.

The magnetic moments induced on the carbon atoms are also position dependent, with the highest induced moment, $+0.08$~$\mu_B$ ($+0.05$~$\mu_B$) for Bz (COT), occurring for the carbon atoms bound directly on top of a Mn atom, i.e. for \Ctopone. However, the total magnetic moment of the organic molecule is small and amounts to only $+0.03$~$\mu_B$, $+0.05$~$\mu_B$, and $0.00$~$\mu_B$ for Bz, COT and BzV, respectively.
Bader charge analysis \cite{Henkelman2006} shows that the Bz (COT) molecule gains 0.77~e (0.96~e) on its adsorption onto the surface. The majority of this charge originates in the four Mn atoms directly underneath the carbon atoms.

For the case of the adsorbed BzV molecule, there is a strong interaction between the V atom and the Mn atoms beneath it. The result is a large change in their respective magnetic moments. While the isolated molecule has a magnetic moment of $1.0$~$\mu_B$ with the majority of this found on the V ion, upon adsorption this is reduced to $-0.3$~$\mu_B$, i.e. directed antiparallel to the moments of the Mn row located below it along the [001] direction (cf.~Fig.\ref{fig:structures}(c)).
The magnitude of these Mn moments is then considerably reduced by $1.81$~$\mu_B$.  The case where the V moment is directed parallel to the Mn row beneath could not be stabilized 
in the calculation. 
Due to the increased distance between the Bz ring and the magnetic surface (3.43~\AA), and the small magnetic moment of the V atom, the moment induced on the Bz layer is negligible.

\subsection{Spin resolved local density of states}\label{ldos}
Despite the small induced magnetic moment on the organic molecules, there is a strong imbalance between the number of spin-up and spin-down states of the molecule in the vicinity of the Fermi level. This is not surprising due to the strong binding of these states with the spin-split states of the Mn atoms beneath the molecules and has been found to be a general feature for all organic molecules adsorbed on ferromagnetic surfaces \cite{PhysRevB.84.172402, PhysRevB.84.224403, Wang2013}.

\begin{figure}
\begin{centering}
\includegraphics[width=\linewidth]{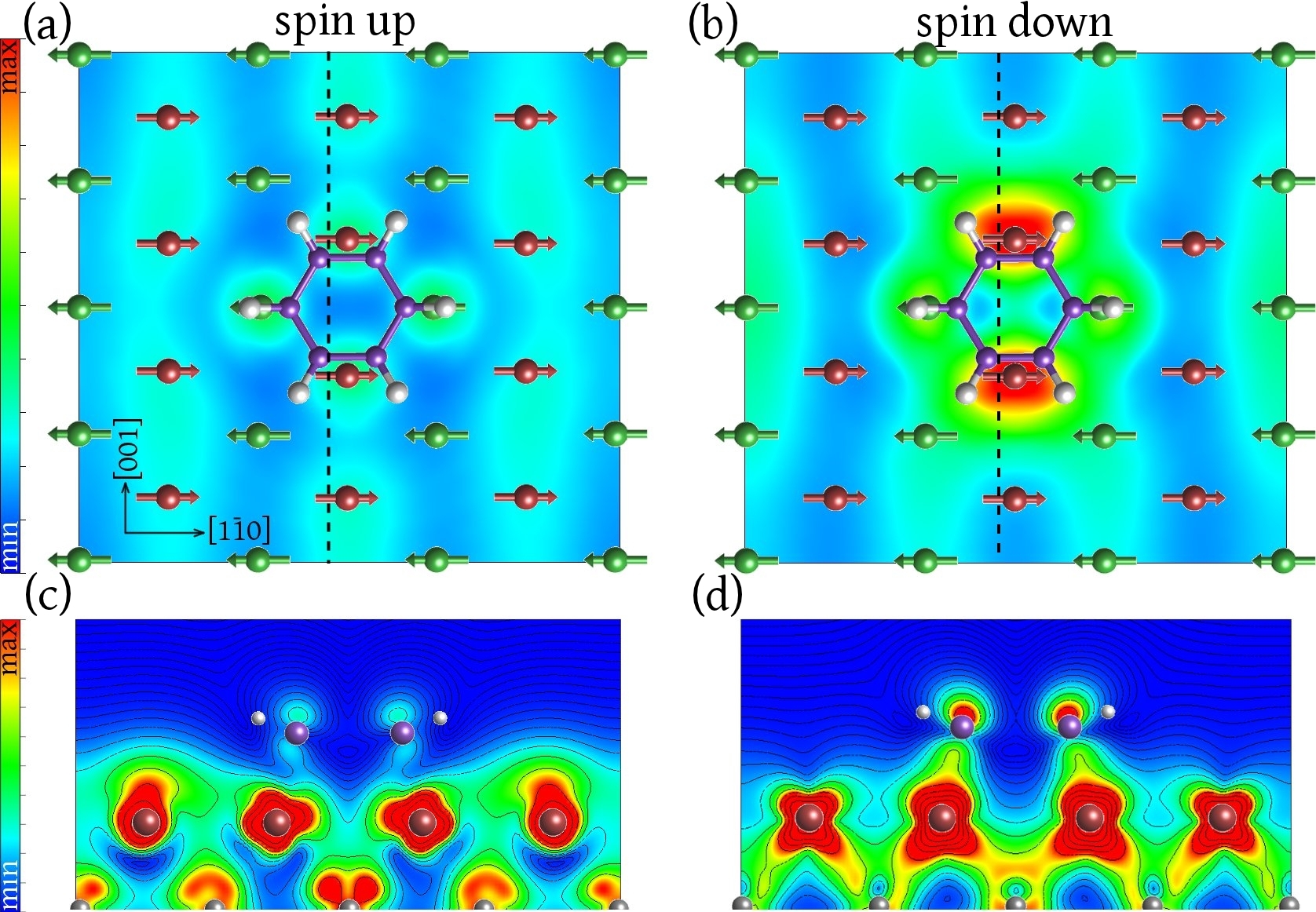}
\caption{\label{fig:bz_spins}(Color online) Charge density in the vacuum 3.0 \AA\ above a Bz molecule adsorbed onto a Mn/W(110) substrate, calculated for occupied energy levels [$-0.4$~eV, $E_{\rm F}$]. (a) shows the spin up contribution to the charge density, (b) the spin down contribution, (c) a slice of the spin up charge density perpendicular to the surface plane along the black dashed line in (a) and (d) a slice of the spin down charge density perpendicular to the surface plane along the black dashed line in (b).}
\end{centering}
\end{figure}

To illustrate this, Fig.~\ref{fig:bz_spins} shows the spin-resolved local density of states integrated in an energy window below the Fermi energy for a Bz molecule adsorbed on a Mn/W(110) surface calculated on a plane parallel to the surface and 5.1 \AA\ above it. Evident for both spin up and spin down charge is the underlying antiferromagnetic pattern of the Mn/W(110) substrate, which is visible as stripes along the [001] direction. As seen in the figure, the Mn row passing directly underneath the molecule has a magnetic moment pointing to the right which in our definition of the SQA means that it has a higher spin down density of states in the vicinity of the Fermi energy.
These states, which have predominately a \dxz\ symmetry, are highly localized on the clean surface and thus suppressed quickly in the vacuum. The spin up states, however, have a spherical $s$-like symmetry. Due to their larger extent, these states decay more slowly in the vacuum and, at a particular height above the surface, dominate despite their lower density at the surface. Clearly, the inverse is also true; Mn atoms with an opposite magnetic moment have a higher spin up density of states which decays quickly in the vacuum, leaving spin down states with an $s$-like symmetry.

This decay of the states into the vacuum can be observed in Figs.~\ref{fig:bz_spins}(c) and (d) which shows cross-sectional plots of the charge density. A comparison of the spin up and spin down channels shows that, far from the influence of the molecule, the spin down charge density at the Mn atoms is indeed larger but that the spin up charge density stretches further into the vacuum.
From Fig.~\ref{fig:bz_spins}, it is clear that the strongest interaction is between the 
$p_z$ orbitals of \Ctoptwo\ and the \dxz\ orbitals of a Mn atom with magnetic moment pointing to the right, resulting in a strong propagation of spin down states into the vacuum. A similar interaction strength could be expected to occur between the \Ctopone\ orbitals and the \dxz\ orbitals of a Mn row with magnetic moment pointing to the left. However, details of the bonding result in a stronger interaction with the $s$-like orbitals.
To summarise, the hybridization of the \pz\ orbitals of the molecule with the $d$-states of the 
surface, results in the enhancement of the spin character of these $d$-orbitals in the vacuum. This will have important repercussions when we consider the vacuum spin polarization in the next section.

\begin{figure}
\begin{centering}
\includegraphics[width=\linewidth]{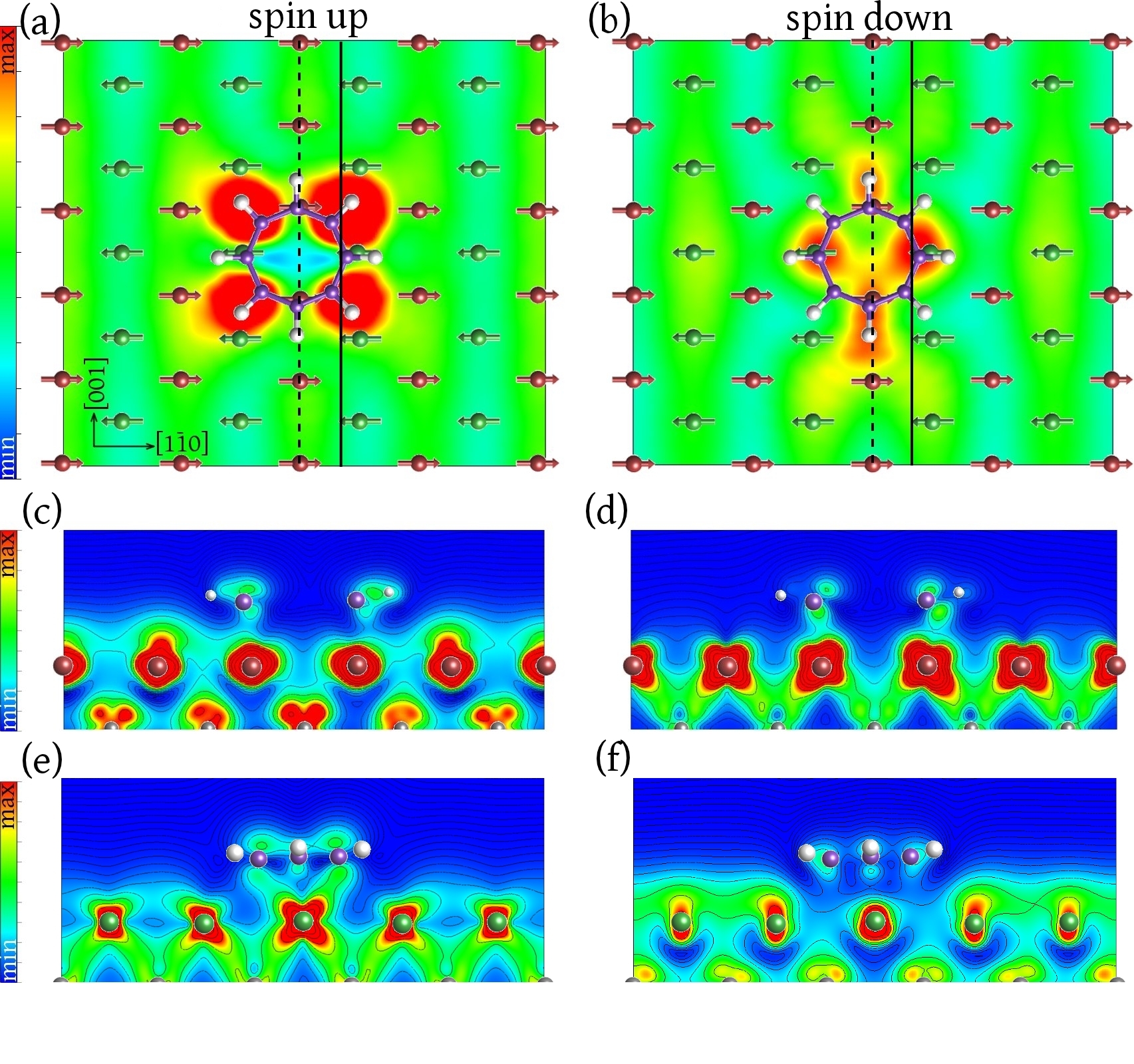}
\caption{\label{fig:cot_spins}(Color online) Charge density in the vacuum 3.0 \AA\ above a COT molecule adsorbed onto a Mn/W(110) substrate, calculated for occupied energy levels [$-0.4$~eV, $E_{\rm F}$]. (a) shows the spin up contribution to the charge density, (b) the spin down contribution, (c) a slice of the spin up charge density perpendicular to the surface plane along the black dashed line in (a), (d) a slice of the spin down charge density along the black dashed line in (b), (e)  a slice of the spin up charge density along the black solid line in (a) and (f) a slice of the spin down charge density along the black solid line in (b).}
\end{centering}
\end{figure}

Fig.~\ref{fig:cot_spins} shows the spin-resolved local density of states integrated in an energy interval of 0.4~eV below the Fermi energy for a COT molecule calculated on the same plane as for the Bz molecule. Here, it is clear that the strongest contribution stems from the spin up states of the four \Cbridge\ atoms. This is due to the hybridization induced broad peak in the DOS of these carbon atoms just below $E_{\rm F}$
(cf. Fig.~\ref{fig:cot_dos}(b)). The spin down charge density in the vacuum is centred around the other four carbon atoms, namely \Ctopone\ and \Ctoptwo. The small asymmetry seen in particular here can be attributed to the slightly asymmetric adsorption position of the COT molecule on the surface.
Fig.~\ref{fig:cot_spins}(c) to (f) show cross sectional plots of the charge density along the [001] direction through the centre of the molecule (c and d) and through the edge of the molecule close to the \Cbridge\ atoms (e and f). One can clearly see the strong hybridization between the carbon \pz\ orbitals and Mn \dxz\ orbitals which was visible in the DOS (cf. Fig.~\ref{fig:cot_dos}), and which leads to the large spin up contribution to the charge density in the vacuum above the \Cbridge\ atoms.

\begin{figure}[hb!]
\begin{centering}
\includegraphics[width=\linewidth]{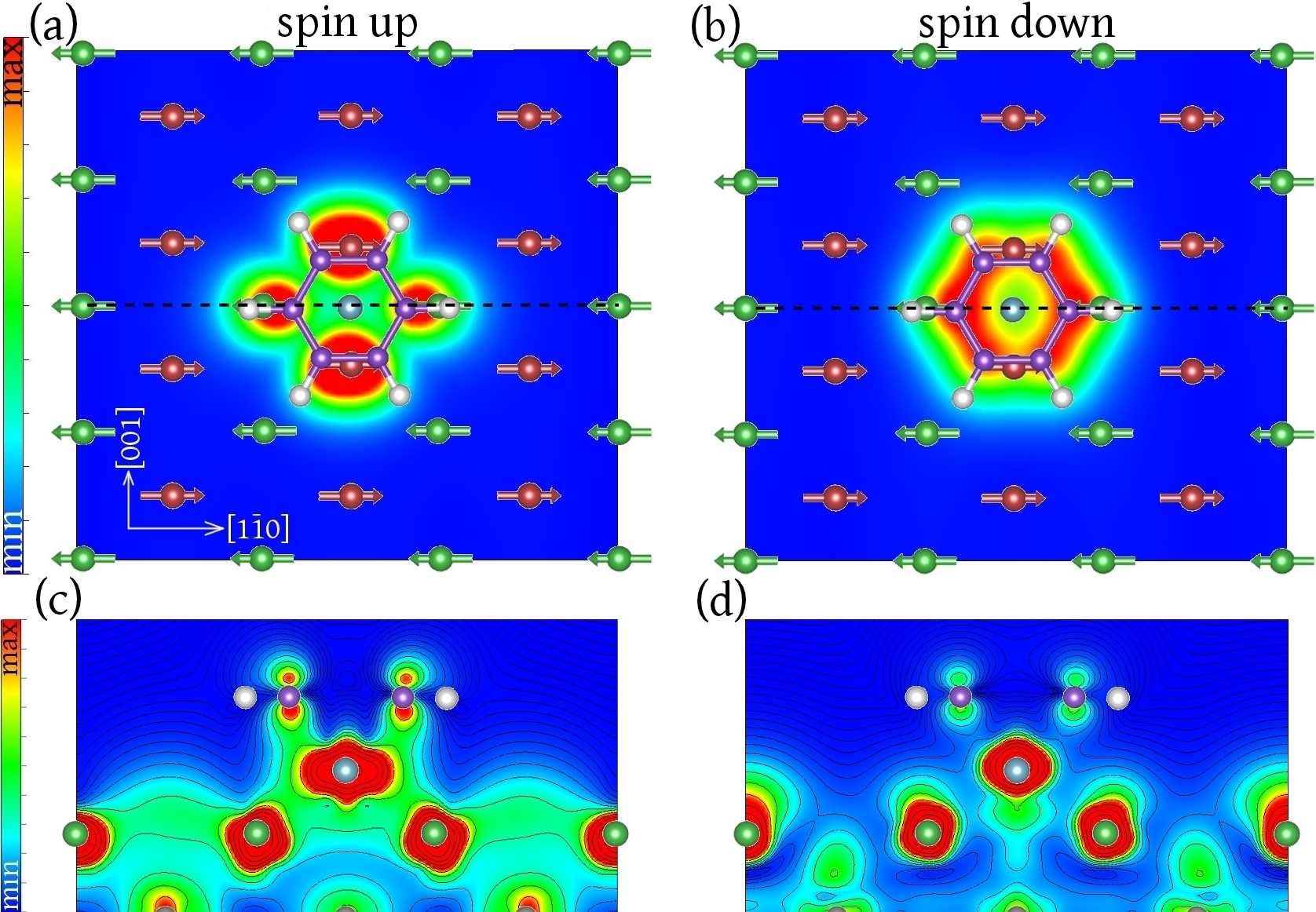}
\caption{\label{fig:bzV_spins}(Color online) Charge density in the vacuum 1.6 \AA\ above a BzV molecule adsorbed onto a Mn/W(110) substrate, calculated for occupied energy levels [$-0.4$~eV, $E_{\rm F}$]. (a) shows the spin up contribution to the charge density, (b) the spin down contribution, (c) a slice of the spin up charge density perpendicular to the surface plane along the black dashed line in (a) and (d)  a slice of the spin down charge density perpendicular to the surface plane along the black dashed line in (b).}
\end{centering}
\end{figure}

Finally, Fig.~\ref{fig:bzV_spins} shows the spin-resolved charge density in the vacuum for the case of an adsorbed BzV molecule, calculated on the same plane as for the Bz and COT molecules, 
i.e.~1.6 \AA\ above the Bz ring \footnote{The shape of the spin resolved charge density in the vacuum, and consequently the vacuum spin polarization, does not change when calculated on a higher plane, e.g. 3~\AA\ above the Bz ring.} . The spin up states of V in this region are dominated by the $d_{x^2 - y^2}$ orbital. The shape of this orbital means that it is bonding approximately equally with both \Ctopone\ and \Ctoptwo\ atoms, 
with little charge density in the bond connecting the two. This hybridization can be seen clearly in the cross-sectional plot in Fig.~\ref{fig:bzV_spins}(c).
In contrast, for spin down, the molecular $\pi$ orbitals of the benzene ring are nearly unperturbed in this energy range.
%

\subsection{Spin polarization in the vacuum}
\begin{figure}[t]
\begin{centering}
\includegraphics[width=\linewidth]{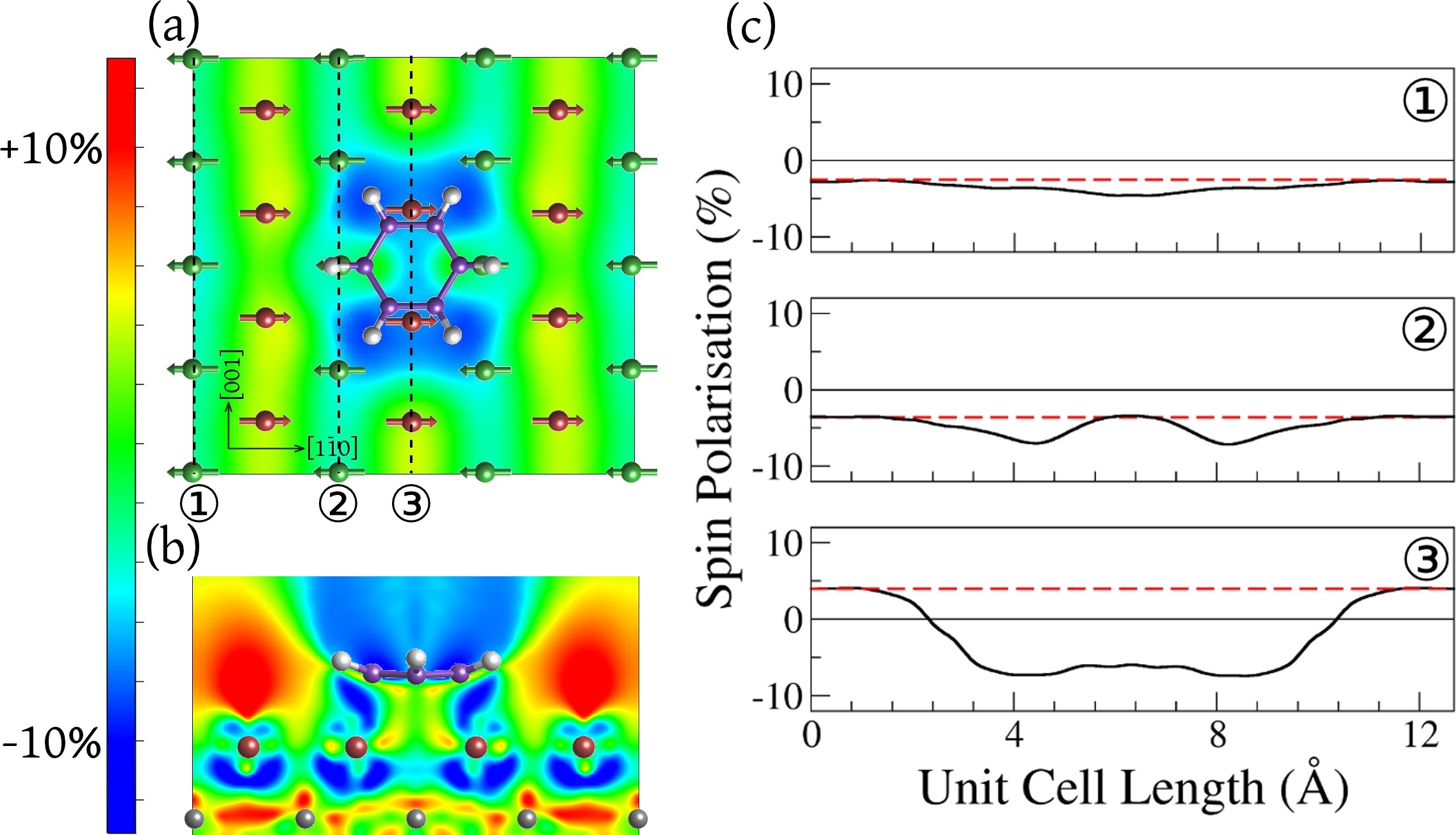}
\caption{\label{fig:bz_spin_polarization}(Color online) (a) Spin polarization 
$P_{\mathrm{vac}}({\mathbf r}_{\scriptscriptstyle \parallel}, z, \epsilon)$
in the vacuum 3.0 \AA\ above a Bz adsorbed onto a Mn/W(110) substrate, calculated for occupied energy levels
[$-0.4$~eV, E$_F$] according to Eq.~(\ref{polarization}). (b) Slice of the spin polarization perpendicular to the surface plane along line \ding{194}. (c) Selected line profiles of the spin polarization. The red dashed line approximates the line profile if no molecule were present.}
\end{centering}
\end{figure}

From the spin resolved charge density in the vacuum we define the spatially resolved spin polarization as:
\begin{equation}\label{polarization}
 P_{\mathrm{vac}}({\mathbf r}_{\scriptscriptstyle \parallel}, z, \epsilon) = \frac{\tilde{n}_s^\uparrow({\mathbf r}_{\scriptscriptstyle \parallel}, z, \epsilon) - \tilde{n}_s^\downarrow({\mathbf r}_{\scriptscriptstyle \parallel} ,z , \epsilon)}{\tilde{n}_s^\uparrow({\mathbf r}_{\scriptscriptstyle \parallel}, z, \epsilon) + \tilde{n}_s^\downarrow({\mathbf r}_{\scriptscriptstyle \parallel}, z, \epsilon)}
\end{equation}
where $\tilde{n}_s^{\uparrow(\downarrow)}({\mathbf r}_{\scriptscriptstyle \parallel},z, \epsilon)$ is the spin up (spin down) charge density calculated
in the vacuum at a lateral position ${\mathbf r}_{\scriptscriptstyle \parallel}$ and a distance, $z$, from the surface, for an energy interval between a chosen energy, $\epsilon$, and the Fermi level ([$\epsilon$, E$_F$]).
This is shown for the case of the Bz molecule in Fig.~\ref{fig:bz_spin_polarization}. Above the molecule there is a strong inversion of spin polarization compared to the bare antiferromagnetic surface.
As discussed in the previous section, when a molecule with $\pi$\ orbitals, such as Bz, is adsorbed on the surface, the \pz\ orbitals couple to the \dxz\ orbitals so that the propagation of this state, whether with positive or negative spin polarization, into the vacuum is extended.
The overall effect is that, when imaged with an STM tip, an inversion of the clean surface spin polarization is observed. The strength of this inversion then depends on the exact details of the bonding.

To illustrate this local inversion of spin polarization, in the right hand panel we present three line profiles through three different rows of Mn atoms. Line \ding{192} shows the spin polarization of a Mn row with magnetic moments to the left far from the influence of the molecule. An average value of $-$3.5\% at a distance of 5.1 \AA\ above the surface is found. Line \ding{193} displays a line profile that cuts through an edge of the molecule. Far from the molecule a background spin polarization of $-$3.5\% is measured. At the edge of the molecule this is increased in magnitude to $-7.0$\% before being reduced again to $-$3.5\% above \Ctopone. Line \ding{194}, cutting through the centre of the molecule, displays a total inversion of spin polarization. Far from the molecule the spin polarization is positive. At a distance of 4.0 \AA\ from the centre of the Bz ring the spin polarization inverts,  i.e.~a distance of 1.86 \AA\ from the physical extent of the molecule in the [001] direction. The spin polarization reaches a maximum value of $-$7.5\% directly above the hydrogen atoms. That is, in addition to an inversion of spin polarization, the magnitude is also enhanced, in this case by a factor of two.

Fig.~\ref{fig:bz_spin_polarization}(b) shows a slice of the spin polarization in a plane perpendicular to that defined by the surface along the [001] direction corresponding to line \ding{194} in Fig.~\ref{fig:bz_spin_polarization}(a). It is clear, that while the spin polarization cutting through the Mn atoms has spin polarization of a particular sign, with a \dxz\ symmetry, the sign is inverted as one extends into the vacuum above the surface.
This effect is due to the slower decay of the $s$-states of the spin up channel as discussed in the previous section.
The adsorption of the molecules clearly disrupts that decay. The hybridization between the \pz\ and \dxz\ states is particularly evident here with the resultant propagation of an inverted spin polarization into the vacuum.

\begin{figure}
\begin{centering}
\includegraphics[width=\linewidth]{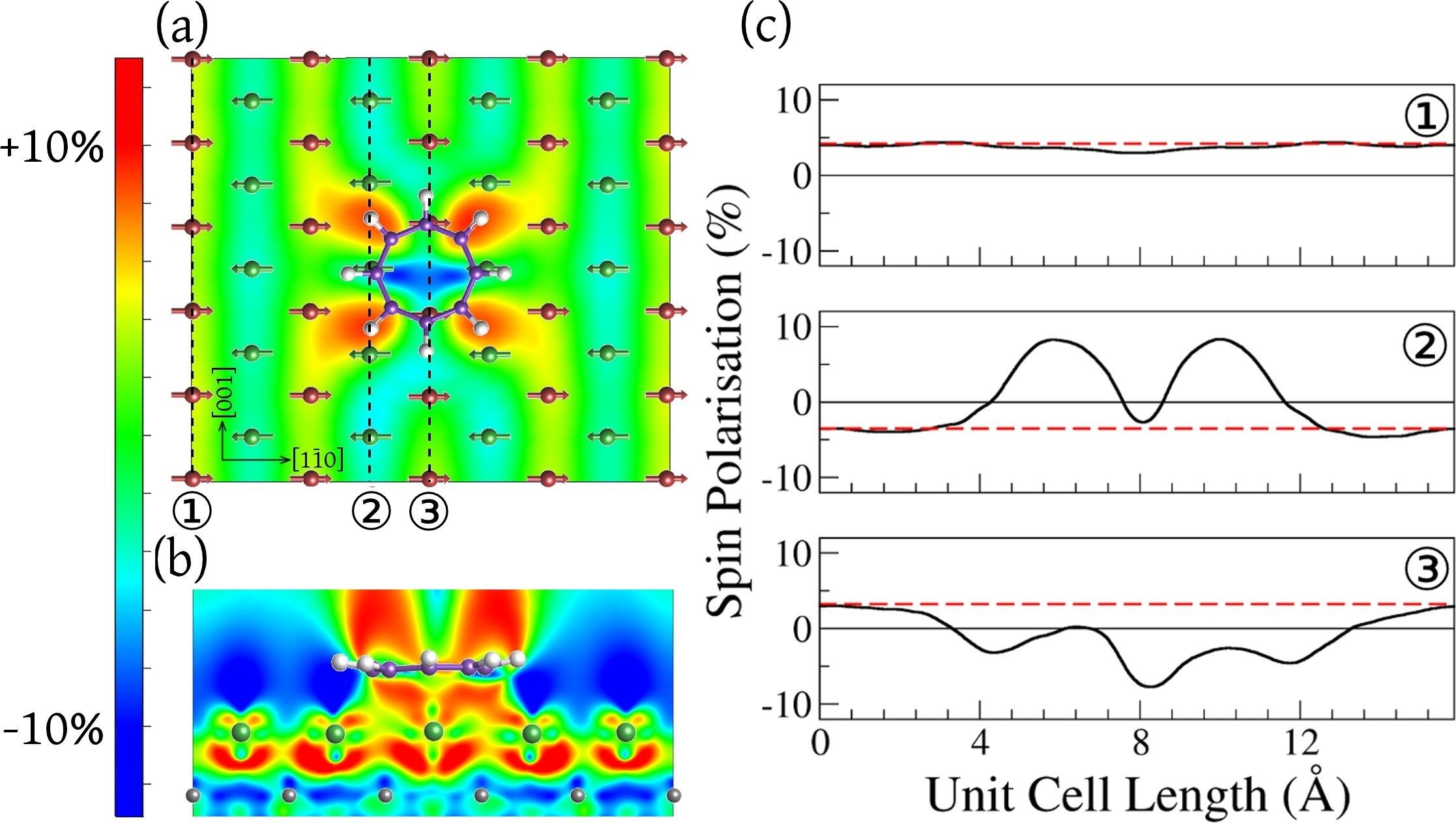}
\caption{\label{fig:cot_spin_polarization}(Color online) (a) Spin polarization 
$P_{\mathrm{vac}}({\mathbf r}_{\scriptscriptstyle \parallel}, z, \epsilon)$
in the vacuum 3.0 \AA\ above a COT adsorbed onto a Mn/W(110) substrate, calculated for occupied energy levels 
[$-0.4$~eV, E$_F$] according to Eq.~(\ref{polarization}). (b) Slice of the spin polarization perpendicular to the surface plane along line \ding{193}. (c) Selected line profiles of the spin polarization. The red dashed line approximates the line profile if no molecule were present.}
\end{centering}
\end{figure}

A similar modulation and inversion of spin polarization occurs in the case of the COT molecule. Due to the asymmetric adsorption and the greater extent of the molecule, however, the effect is more pronounced. This can be seen in Fig.~\ref{fig:cot_spin_polarization}. Far from the COT molecule, line \ding{192} shows a stable positive spin polarization of approximately $+$3.5\%. By line \ding{193} a dramatic inversion can be observed, going from $-$3.5\% far from the molecule to $+$8.2\% above the \Cbridge\ atoms. Similarly, line \ding{194} shows a large inversion, although in the opposite direction; from positive values far from the molecule to $-$8.2\% directly in the centre of the COT ring, an enhancement of over 100\%. Note that the adsorption asymmetry can also be easily detected in the line profile.

\begin{figure}
\begin{centering}
\includegraphics[width=\linewidth]{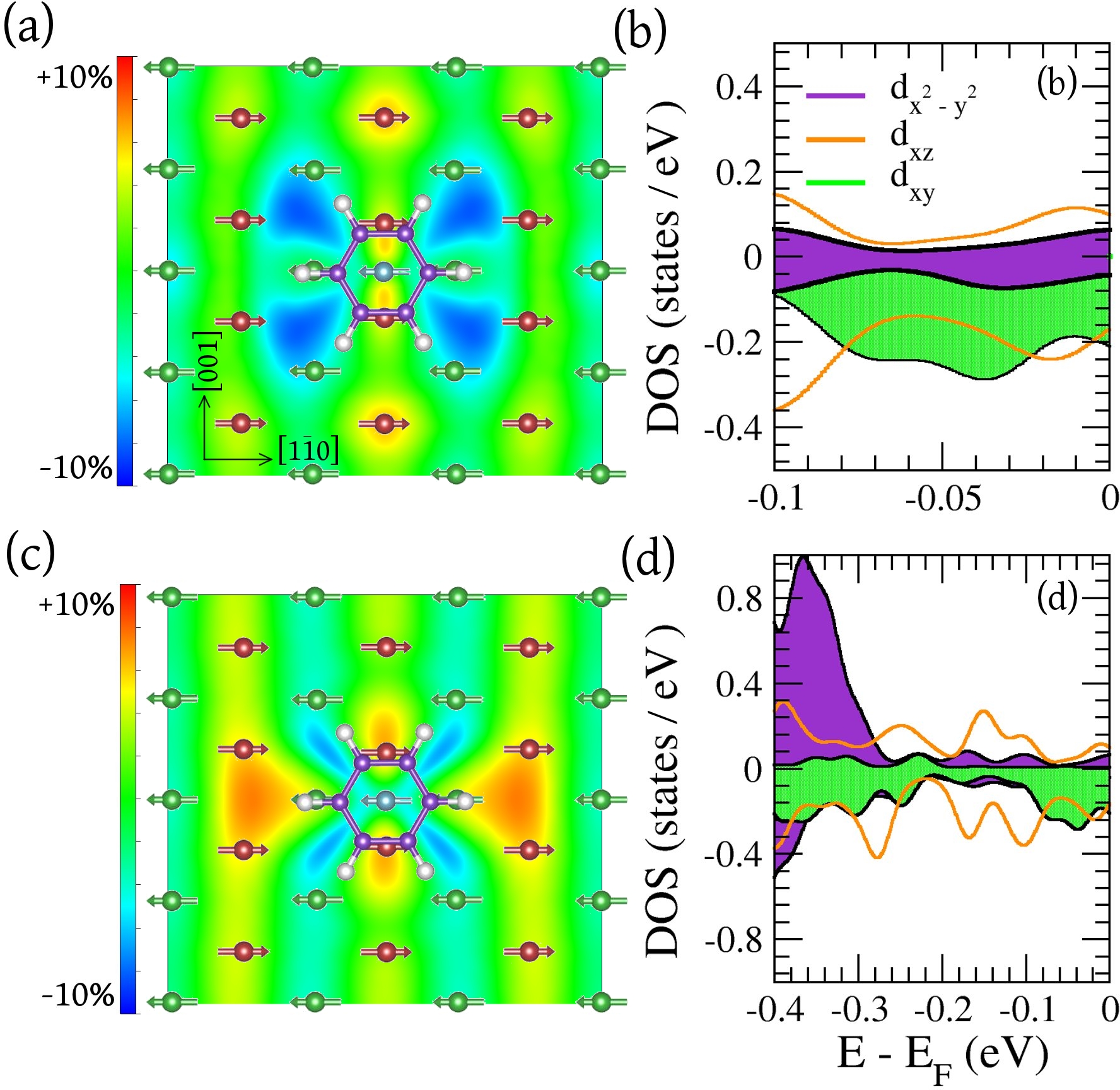}
\caption{\label{fig:bzV_sp_and_dos}(Color online) Spin polarization in the vacuum 1.6 \AA\ above a BzV molecule adsorbed onto a Mn/W(110) substrate, calculated for occupied energy levels 
[$-0.1$~eV, $E_{\rm F}$] ((a) and (b)) [$-0.4$~eV, $E_{\rm F}$] ((c) and (d))  according to Eq.~(\ref{polarization}). The spin resolved density of states are shown for the most relevant $d$ orbitals of the adsorbed V atom.}
\end{centering}
\end{figure}

An examination of the density of states close to the Fermi level of the carbon atoms in Figs.~\ref{fig:bz_dos}, \ref{fig:cot_dos} and \ref{fig:bzV_dos} suggests that the spin polarization could be quite energy dependent, varying considerably for different energy ranges. As an example case, we look at the spin polarization in the vacuum for a BzV molecule adsorbed on the surface.
As we have seen, the spin polarization in the vacuum above the adsorbed molecule is driven by the binding between the molecule and the magnetic surface atoms.
The change of polarization in the vacuum above the BzV molecule is induced now, not by the direct interaction of the $\pi$ orbitals with the Mn surface, but indirectly via the $d$ orbitals of the V atom close to the Fermi level.
This is demonstrated in Fig.~\ref{fig:bzV_sp_and_dos}. Panels (a) and (b) show the spin polarization in the vacuum and LDOS of the V atom in an energy window between [$-0.1$~eV, E$_F$]. It is evident that the higher spin down density of states of the V atom results in the four negative lobes of spin polarization above the molecule. The observed cross-like symmetry is due to the hybridization in the spin up channel between the V \dxz\ orbitals and the Bz $\pi$ orbitals while in the spin down channel it is the hybridization between the in-plane V \dxy\ orbital and the carbon \pz\ orbitals that plays a more important role.

Fig.~\ref{fig:bzV_sp_and_dos}(c) and (d) show the effect of increasing the size of the energy window probed to [$-0.4$~eV, E$_F$]. Here it is clear that the $d_{x^2 - y^2}$ orbital is the dominant contribution to the the spin up states. The hybridization of this orbital and the \Ctopone\ atoms is very strong and results in the large area of positive spin polarization located approximately above the \Ctopone\ atoms and extending beyond the physical extent of the molecule.
It is remarkable that due to the adsorbed Bz ring atomic $d$-states of a symmetry type which are normally not detectable by
STM, such as \dxz\ or $d_{x^2 - y^2}$, play a major role for the spin polarization and image of a transition-metal-benzene complex.

\subsection{Modelling the effect of the spin spiral}

Finally, we consider how the experimentally observed spin spiral ground state of Mn/W(110) might influence the measured SP-STM images. As stated in the introduction, the monolayer of Mn grown pseudomorphically on a W(110) substrate
exhibits a spin spiral which propagates along the [$1\bar{1}0$] direction with a small angle of $\sim$173$^\circ$ between moments on adjacent rows. Although the orientation direction relative to the surface plane changes quite slowly, across the width of the 13.45 \AA\ unit cell considered here there is a change of angle of 45$^\circ$ and so it is worth considering.

The tunnelingcurrent between the sample (S) and the tip (T) can be described by the spin-polarized Tersoff-Hamann 
model \cite{PhysRevLett.86.4132} and written in the following form:
\begin{multline}\label{rho_spiral}
I(\vec{r}, \Theta)\propto \nicefrac{1}{2} [\:(\:1 + P_\mathrm{T}\cos\Theta(\vec{r}))\:n_\mathrm{s}^\uparrow (\vec{r})\: + \\ (\:1 - P_\mathrm{T}\cos\Theta(\vec{r}))\: n_\mathrm{s}^\downarrow (\vec{r}\:)\:]
\end{multline}
where the spin polarization of the tip is given by $P_{\mathrm{T}} = (n_\mathrm{T}^\uparrow - n_\mathrm{T}^\downarrow)/(n_\mathrm{T}^\uparrow + n_\mathrm{T}^\downarrow)$, $\Theta$ is the angle between the tip and local sample magnetization at position $\vec{r}$ and $n_\mathrm{s}^\uparrow(\downarrow)$ is the spin up (spin down) charge density of the surface.

\begin{figure}[h!]
\begin{centering}
\includegraphics[width=0.95\linewidth]{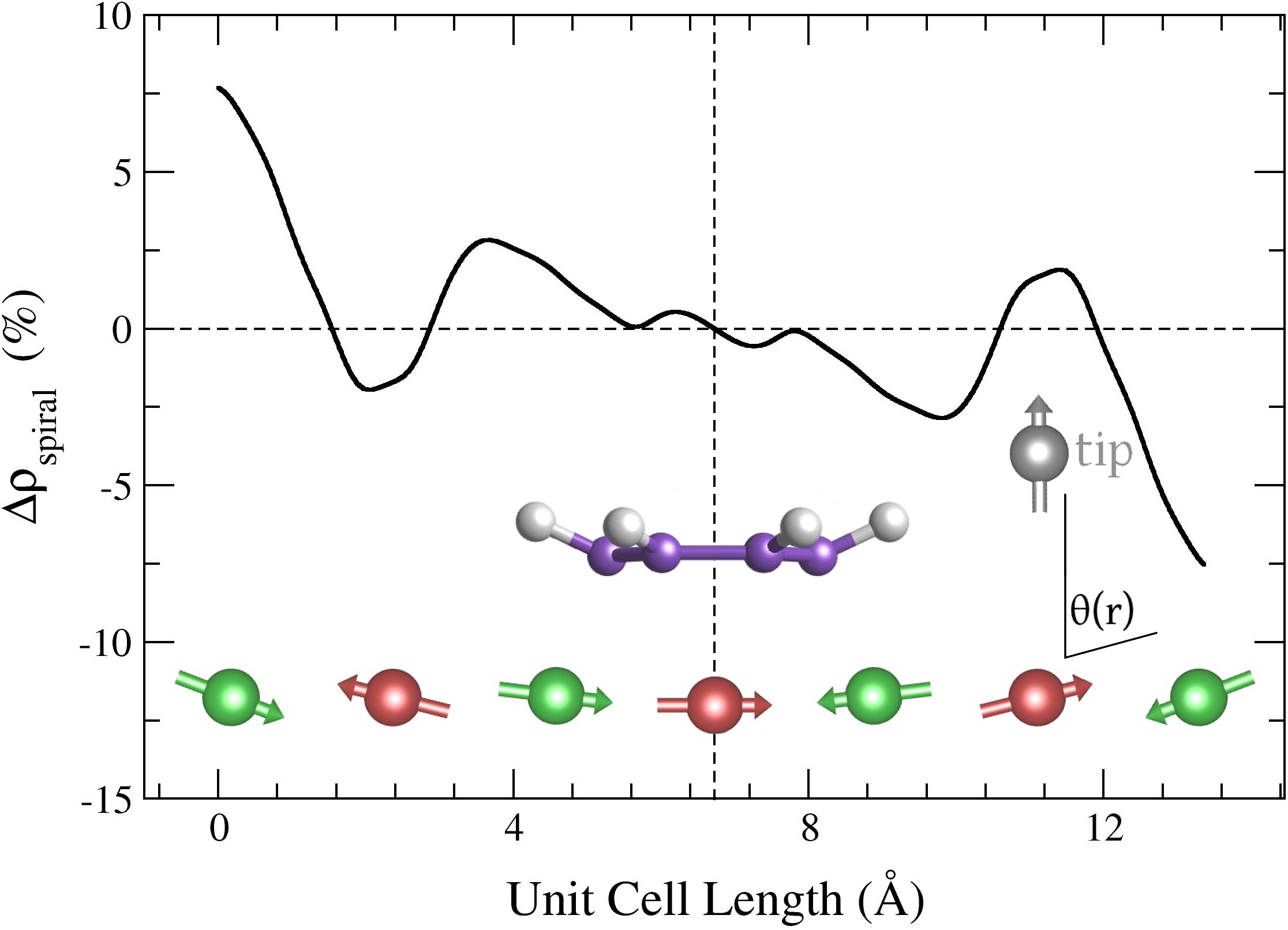}
\caption{\label{fig:spiral}(Color online) $\Delta \rho_{\mathrm{spiral}}$, as defined in Eq.~\ref{rho_spiral}, calculated in the vacuum 1.6~\AA\ above an Bz molecule adsorbed onto a Mn/W(110) substrate, for the occupied energy levels 
[$-0.4$~eV,  $E_{\rm F}$].}
\end{centering}
\end{figure}

The strength of the influence of the spin spiral on the simulated SP-STM images in a
[$-0.4$~eV, E$_F$] energy region, can be seen in Fig.~\ref{fig:spiral} for the example case of an adsorbed Bz molecule. The line profile is drawn at the same height above the surface as in Fig.~\ref{fig:bz_spin_polarization}. Here, we have assumed that the angle between the tip magnetization and the row of Mn atoms at the centre of the unit cell is 90$^\circ$. At the left hand edge of the unit cell, the angle between the tip and sample increases to 111$^\circ$, while the right-most row of Mn atoms have an angle of 69$^\circ$ between tip and surface. A tip polarization, $P_{\mathrm{T}}$, of 70\% is assumed. Fig.~\ref{fig:spiral} shows the difference between the charge density in the vacuum as determined by the spin polarized tip and that with a non-magnetic tip (or equivalently, by assuming a constant 90$^\circ$ angle between the magnetization of the tip and sample across the whole unit cell), i.e.,
$$
\Delta \rho_{\mathrm{spiral}} = \frac{I(\vec{r}, \Theta) - I(\vec{r}, 90^\circ)}{I(\vec{r}, 90^\circ)}
$$
The line profile shown is drawn through the centre of the molecule along the [$1\bar{1}0$] direction of propagation of the spin spiral.
By construction, $\Delta \rho_{\mathrm{spiral}}$ is zero at the centre of the cell. It can be seen that the maximum asymmetry induced by the spin spiral across the molecule is $\approx$~6\%, from $+3$\% above the edge of one side of the molecule to $-3$\% above the other side. 
While our simple model captures the main effect of an asymmetric spin-polarized STM contrast it does not explicitly contain the
non-collinear spin structure of the surface and therefore cannot take into account changes in the electronic structure at the
interface or induced by spin-orbit coupling. 
Nonetheless, the minor asymmetry induced by the spin spiral validates our use of the collinear approximation.

\section{Summary and Conclusions}
In this work, we have demonstrated that the spin polarization of a hybrid organic - inorganic interface can have a strong spatial dependence varying on the scale of the adsorbed molecule and with a magnitude and sign that change relative to that which would be measured on a clean surface. This effect occurs for magnetic surfaces with a spin structure which varies on the atomic scale such as the Mn monolayer on W(110) even for small organic molecules such as Bz, COT or BzV.  Despite the negligible magnetic moment induced on the adsorbents, the spin polarization in the vacuum was found to be on the order of 10\% with different regions of the molecule capable of sustaining different signs of spin polarization. This effect is shown to be driven by the hybridization of molecular orbitals with the spin split orbitals of the surface. For the clean surface, it is states with $s$-like symmetry that dominate in the vacuum despite their low density due to the fast decay of states with $d$ character. The adsorption of an organic molecule disrupts this decay. The hybridization of the $p$ states of the molecule with the $d$ states of the metal is shown to result in the extension of the $d$-states spin character, whether spin up or spin down, into the vacuum.

\begin{acknowledgments}
This work has been supported by Deutsche Forschungsgemeinschaft via the project B10 of the SFB 677 and FCRM ``THE-SIMS''. Computational facilities were provided by the North-German Supercomputing Alliance (HLRN). The authors would like to thank Valerio Bellini for useful discussions. The structures and the 2D graphs of the charge density were produced using the VESTA software \cite{vesta}.
\end{acknowledgments}


\begin{thebibliography}{57}
\expandafter\ifx\csname natexlab\endcsname\relax\def\natexlab#1{#1}\fi
\expandafter\ifx\csname bibnamefont\endcsname\relax
  \def\bibnamefont#1{#1}\fi
\expandafter\ifx\csname bibfnamefont\endcsname\relax
  \def\bibfnamefont#1{#1}\fi
\expandafter\ifx\csname citenamefont\endcsname\relax
  \def\citenamefont#1{#1}\fi
\expandafter\ifx\csname url\endcsname\relax
  \def\url#1{\texttt{#1}}\fi
\expandafter\ifx\csname urlprefix\endcsname\relax\def\urlprefix{URL }\fi
\providecommand{\bibinfo}[2]{#2}
\providecommand{\eprint}[2][]{\url{#2}}

\bibitem[{\citenamefont{Bogani and Wernsdorfer}(2008)}]{BoganiReview2008}
\bibinfo{author}{\bibfnamefont{L.}~\bibnamefont{Bogani}} \bibnamefont{and}
  \bibinfo{author}{\bibfnamefont{W.}~\bibnamefont{Wernsdorfer}},
  \bibinfo{journal}{Nature Materials} \textbf{\bibinfo{volume}{7}},
  \bibinfo{pages}{179} (\bibinfo{year}{2008}).

\bibitem[{\citenamefont{Dediu et~al.}(2009)\citenamefont{Dediu, Hueso,
  Bergenti, and Taliani}}]{Dediu2009}
\bibinfo{author}{\bibfnamefont{V.~A.} \bibnamefont{Dediu}},
  \bibinfo{author}{\bibfnamefont{L.~E.} \bibnamefont{Hueso}},
  \bibinfo{author}{\bibfnamefont{I.}~\bibnamefont{Bergenti}}, \bibnamefont{and}
  \bibinfo{author}{\bibfnamefont{C.}~\bibnamefont{Taliani}},
  \bibinfo{journal}{Nature Materials} \textbf{\bibinfo{volume}{8}},
  \bibinfo{pages}{707} (\bibinfo{year}{2009}).

\bibitem[{\citenamefont{Sanvito}(2011)}]{SanvitoReview2011}
\bibinfo{author}{\bibfnamefont{S.}~\bibnamefont{Sanvito}},
  \bibinfo{journal}{Chemical Society Reviews} \textbf{\bibinfo{volume}{40}},
  \bibinfo{pages}{3336} (\bibinfo{year}{2011}).

\bibitem[{\citenamefont{Dediu et~al.}(2002)\citenamefont{Dediu, Murgia,
  Matacotta, Taliani, and Barbanera}}]{Dediu2002}
\bibinfo{author}{\bibfnamefont{V.}~\bibnamefont{Dediu}},
  \bibinfo{author}{\bibfnamefont{M.}~\bibnamefont{Murgia}},
  \bibinfo{author}{\bibfnamefont{F.}~\bibnamefont{Matacotta}},
  \bibinfo{author}{\bibfnamefont{C.}~\bibnamefont{Taliani}}, \bibnamefont{and}
  \bibinfo{author}{\bibfnamefont{S.}~\bibnamefont{Barbanera}},
  \bibinfo{journal}{Solid State Communications} \textbf{\bibinfo{volume}{122}},
  \bibinfo{pages}{181} (\bibinfo{year}{2002}).

\bibitem[{\citenamefont{Pramanik et~al.}({2007})\citenamefont{Pramanik,
  Stefanita, Patibandla, Bandyopadhyay, Garre, Harth, and
  Cahay}}]{Pramanik2007}
\bibinfo{author}{\bibfnamefont{S.}~\bibnamefont{Pramanik}},
  \bibinfo{author}{\bibfnamefont{C.~G.} \bibnamefont{Stefanita}},
  \bibinfo{author}{\bibfnamefont{S.}~\bibnamefont{Patibandla}},
  \bibinfo{author}{\bibfnamefont{S.}~\bibnamefont{Bandyopadhyay}},
  \bibinfo{author}{\bibfnamefont{K.}~\bibnamefont{Garre}},
  \bibinfo{author}{\bibfnamefont{N.}~\bibnamefont{Harth}}, \bibnamefont{and}
  \bibinfo{author}{\bibfnamefont{M.}~\bibnamefont{Cahay}},
  \bibinfo{journal}{{Nature Nanotechnology}} \textbf{\bibinfo{volume}{{2}}},
  \bibinfo{pages}{{216}} (\bibinfo{year}{{2007}}).

\bibitem[{\citenamefont{Xiong et~al.}({2004})\citenamefont{Xiong, Wu, Vardeny,
  and Shi}}]{Xiong2004}
\bibinfo{author}{\bibfnamefont{Z.}~\bibnamefont{Xiong}},
  \bibinfo{author}{\bibfnamefont{D.}~\bibnamefont{Wu}},
  \bibinfo{author}{\bibfnamefont{Z.}~\bibnamefont{Vardeny}}, \bibnamefont{and}
  \bibinfo{author}{\bibfnamefont{J.}~\bibnamefont{Shi}},
  \bibinfo{journal}{{Nature}} \textbf{\bibinfo{volume}{{427}}},
  \bibinfo{pages}{{821}} (\bibinfo{year}{{2004}}).

\bibitem[{\citenamefont{Barraud et~al.}({2010})\citenamefont{Barraud, Seneor,
  Mattana, Fusil, Bouzehouane, Deranlot, Graziosi, Hueso, Bergenti, Dediu
  et~al.}}]{Barraud2010}
\bibinfo{author}{\bibfnamefont{C.}~\bibnamefont{Barraud}},
  \bibinfo{author}{\bibfnamefont{P.}~\bibnamefont{Seneor}},
  \bibinfo{author}{\bibfnamefont{R.}~\bibnamefont{Mattana}},
  \bibinfo{author}{\bibfnamefont{S.}~\bibnamefont{Fusil}},
  \bibinfo{author}{\bibfnamefont{K.}~\bibnamefont{Bouzehouane}},
  \bibinfo{author}{\bibfnamefont{C.}~\bibnamefont{Deranlot}},
  \bibinfo{author}{\bibfnamefont{P.}~\bibnamefont{Graziosi}},
  \bibinfo{author}{\bibfnamefont{L.}~\bibnamefont{Hueso}},
  \bibinfo{author}{\bibfnamefont{I.}~\bibnamefont{Bergenti}},
  \bibinfo{author}{\bibfnamefont{V.}~\bibnamefont{Dediu}},
  \bibnamefont{et~al.}, \bibinfo{journal}{{Nature Physics}}
  \textbf{\bibinfo{volume}{{6}}}, \bibinfo{pages}{{615}}
  (\bibinfo{year}{{2010}}).

\bibitem[{\citenamefont{Rocha et~al.}({2005})\citenamefont{Rocha,
  Garcia-Suarez, Bailey, Lambert, Ferrer, and Sanvito}}]{Rocha2005}
\bibinfo{author}{\bibfnamefont{A.}~\bibnamefont{Rocha}},
  \bibinfo{author}{\bibfnamefont{V.}~\bibnamefont{Garcia-Suarez}},
  \bibinfo{author}{\bibfnamefont{S.}~\bibnamefont{Bailey}},
  \bibinfo{author}{\bibfnamefont{C.}~\bibnamefont{Lambert}},
  \bibinfo{author}{\bibfnamefont{J.}~\bibnamefont{Ferrer}}, \bibnamefont{and}
  \bibinfo{author}{\bibfnamefont{S.}~\bibnamefont{Sanvito}},
  \bibinfo{journal}{{Nature Materials}} \textbf{\bibinfo{volume}{{4}}},
  \bibinfo{pages}{{335}} (\bibinfo{year}{{2005}}).

\bibitem[{\citenamefont{Haiss et~al.}({2006})\citenamefont{Haiss, Wang, Grace,
  Batsanov, Schiffrin, Higgins, Bryce, Lambert, and Nichols}}]{Haiss2006}
\bibinfo{author}{\bibfnamefont{W.}~\bibnamefont{Haiss}},
  \bibinfo{author}{\bibfnamefont{C.}~\bibnamefont{Wang}},
  \bibinfo{author}{\bibfnamefont{I.}~\bibnamefont{Grace}},
  \bibinfo{author}{\bibfnamefont{A.~S.} \bibnamefont{Batsanov}},
  \bibinfo{author}{\bibfnamefont{D.~J.} \bibnamefont{Schiffrin}},
  \bibinfo{author}{\bibfnamefont{S.~J.} \bibnamefont{Higgins}},
  \bibinfo{author}{\bibfnamefont{M.~R.} \bibnamefont{Bryce}},
  \bibinfo{author}{\bibfnamefont{C.~J.} \bibnamefont{Lambert}},
  \bibnamefont{and} \bibinfo{author}{\bibfnamefont{R.~J.}
  \bibnamefont{Nichols}}, \bibinfo{journal}{{Nature Materials}}
  \textbf{\bibinfo{volume}{{5}}}, \bibinfo{pages}{{995}}
  (\bibinfo{year}{{2006}}).

\bibitem[{\citenamefont{Tao}({2006})}]{Tao2006}
\bibinfo{author}{\bibfnamefont{N.~J.} \bibnamefont{Tao}},
  \bibinfo{journal}{{Nature Nanotechnology}} \textbf{\bibinfo{volume}{{1}}},
  \bibinfo{pages}{{173}} (\bibinfo{year}{{2006}}).

\bibitem[{\citenamefont{Bogani and Wernsdorfer}({2008})}]{Bogani2008}
\bibinfo{author}{\bibfnamefont{L.}~\bibnamefont{Bogani}} \bibnamefont{and}
  \bibinfo{author}{\bibfnamefont{W.}~\bibnamefont{Wernsdorfer}},
  \bibinfo{journal}{{Nature Materials}} \textbf{\bibinfo{volume}{{7}}},
  \bibinfo{pages}{{179}} (\bibinfo{year}{{2008}}).

\bibitem[{\citenamefont{Sanvito and Rocha}({2006})}]{Sanvito2006}
\bibinfo{author}{\bibfnamefont{S.}~\bibnamefont{Sanvito}} \bibnamefont{and}
  \bibinfo{author}{\bibfnamefont{A.~R.} \bibnamefont{Rocha}},
  \bibinfo{journal}{{Journal of Computational and Theoretical Nanoscience}}
  \textbf{\bibinfo{volume}{{3}}}, \bibinfo{pages}{{624}}
  (\bibinfo{year}{{2006}}).

\bibitem[{\citenamefont{Waldron et~al.}(2006)\citenamefont{Waldron, Haney,
  Larade, MacDonald, and Guo}}]{PhysRevLett.96.166804}
\bibinfo{author}{\bibfnamefont{D.}~\bibnamefont{Waldron}},
  \bibinfo{author}{\bibfnamefont{P.}~\bibnamefont{Haney}},
  \bibinfo{author}{\bibfnamefont{B.}~\bibnamefont{Larade}},
  \bibinfo{author}{\bibfnamefont{A.}~\bibnamefont{MacDonald}},
  \bibnamefont{and} \bibinfo{author}{\bibfnamefont{H.}~\bibnamefont{Guo}},
  \bibinfo{journal}{Phys. Rev. Lett.} \textbf{\bibinfo{volume}{96}},
  \bibinfo{pages}{166804} (\bibinfo{year}{2006}).

\bibitem[{\citenamefont{Senapati et~al.}(2007)\citenamefont{Senapati, Pati, and
  Erwin}}]{PhysRevB.76.024438}
\bibinfo{author}{\bibfnamefont{L.}~\bibnamefont{Senapati}},
  \bibinfo{author}{\bibfnamefont{R.}~\bibnamefont{Pati}}, \bibnamefont{and}
  \bibinfo{author}{\bibfnamefont{S.~C.} \bibnamefont{Erwin}},
  \bibinfo{journal}{Phys. Rev. B} \textbf{\bibinfo{volume}{76}},
  \bibinfo{pages}{024438} (\bibinfo{year}{2007}).

\bibitem[{\citenamefont{Rocha and Sanvito}(2007)}]{Rocha2007}
\bibinfo{author}{\bibfnamefont{A.~R.} \bibnamefont{Rocha}} \bibnamefont{and}
  \bibinfo{author}{\bibfnamefont{S.}~\bibnamefont{Sanvito}},
  \bibinfo{journal}{Journal of Applied Physics} \textbf{\bibinfo{volume}{101}},
  \bibinfo{pages}{09B102} (\bibinfo{year}{2007}).

\bibitem[{\citenamefont{Saffarzadeh}({2008})}]{Saffarzadeh2008}
\bibinfo{author}{\bibfnamefont{A.}~\bibnamefont{Saffarzadeh}},
  \bibinfo{journal}{{Journal of Applied Physics}}
  \textbf{\bibinfo{volume}{{104}}}, \bibinfo{pages}{{123715}}
  (\bibinfo{year}{{2008}}).

\bibitem[{\citenamefont{Schmaus et~al.}({2011})\citenamefont{Schmaus, Bagrets,
  Nahas, Yamada, Bork, Bowen, Beaurepaire, Evers, and Wulfhekel}}]{Schmaus2011}
\bibinfo{author}{\bibfnamefont{S.}~\bibnamefont{Schmaus}},
  \bibinfo{author}{\bibfnamefont{A.}~\bibnamefont{Bagrets}},
  \bibinfo{author}{\bibfnamefont{Y.}~\bibnamefont{Nahas}},
  \bibinfo{author}{\bibfnamefont{T.~K.} \bibnamefont{Yamada}},
  \bibinfo{author}{\bibfnamefont{A.}~\bibnamefont{Bork}},
  \bibinfo{author}{\bibfnamefont{M.}~\bibnamefont{Bowen}},
  \bibinfo{author}{\bibfnamefont{E.}~\bibnamefont{Beaurepaire}},
  \bibinfo{author}{\bibfnamefont{F.}~\bibnamefont{Evers}}, \bibnamefont{and}
  \bibinfo{author}{\bibfnamefont{W.}~\bibnamefont{Wulfhekel}},
  \bibinfo{journal}{{Nature Nanotech.}} \textbf{\bibinfo{volume}{{6}}},
  \bibinfo{pages}{{185}} (\bibinfo{year}{{2011}}).

\bibitem[{\citenamefont{Urdampilleta et~al.}({2011})\citenamefont{Urdampilleta,
  Klyatskaya, Cleuziou, Ruben, and Wernsdorfer}}]{Urdampilleta2011}
\bibinfo{author}{\bibfnamefont{M.}~\bibnamefont{Urdampilleta}},
  \bibinfo{author}{\bibfnamefont{S.}~\bibnamefont{Klyatskaya}},
  \bibinfo{author}{\bibfnamefont{J.-P.} \bibnamefont{Cleuziou}},
  \bibinfo{author}{\bibfnamefont{M.}~\bibnamefont{Ruben}}, \bibnamefont{and}
  \bibinfo{author}{\bibfnamefont{W.}~\bibnamefont{Wernsdorfer}},
  \bibinfo{journal}{{Nature Materials}} \textbf{\bibinfo{volume}{{10}}},
  \bibinfo{pages}{{502}} (\bibinfo{year}{{2011}}).

\bibitem[{\citenamefont{Sanvito}({2010})}]{Sanvito2010}
\bibinfo{author}{\bibfnamefont{S.}~\bibnamefont{Sanvito}},
  \bibinfo{journal}{{Nature Physics}} \textbf{\bibinfo{volume}{{6}}},
  \bibinfo{pages}{{562}} (\bibinfo{year}{{2010}}).

\bibitem[{\citenamefont{Kawahara et~al.}({2012})\citenamefont{Kawahara,
  Lagoute, Repain, Chacon, Girard, Rousset, Smogunov, and
  Barreteau}}]{Kawahara2012}
\bibinfo{author}{\bibfnamefont{S.~L.} \bibnamefont{Kawahara}},
  \bibinfo{author}{\bibfnamefont{J.}~\bibnamefont{Lagoute}},
  \bibinfo{author}{\bibfnamefont{V.}~\bibnamefont{Repain}},
  \bibinfo{author}{\bibfnamefont{C.}~\bibnamefont{Chacon}},
  \bibinfo{author}{\bibfnamefont{Y.}~\bibnamefont{Girard}},
  \bibinfo{author}{\bibfnamefont{S.}~\bibnamefont{Rousset}},
  \bibinfo{author}{\bibfnamefont{A.}~\bibnamefont{Smogunov}}, \bibnamefont{and}
  \bibinfo{author}{\bibfnamefont{C.}~\bibnamefont{Barreteau}},
  \bibinfo{journal}{{Nano Letters}} \textbf{\bibinfo{volume}{{12}}},
  \bibinfo{pages}{{4558}} (\bibinfo{year}{{2012}}).

\bibitem[{\citenamefont{Schw\"obel et~al.}({2012})\citenamefont{Schw\"obel, Fu,
  Brede, Dilullo, Hoffmann, Klyatskaya, Ruben, and
  Wiesendanger}}]{Schwoebel2012}
\bibinfo{author}{\bibfnamefont{J.}~\bibnamefont{Schw\"obel}},
  \bibinfo{author}{\bibfnamefont{Y.}~\bibnamefont{Fu}},
  \bibinfo{author}{\bibfnamefont{J.}~\bibnamefont{Brede}},
  \bibinfo{author}{\bibfnamefont{A.}~\bibnamefont{Dilullo}},
  \bibinfo{author}{\bibfnamefont{G.}~\bibnamefont{Hoffmann}},
  \bibinfo{author}{\bibfnamefont{S.}~\bibnamefont{Klyatskaya}},
  \bibinfo{author}{\bibfnamefont{M.}~\bibnamefont{Ruben}}, \bibnamefont{and}
  \bibinfo{author}{\bibfnamefont{R.}~\bibnamefont{Wiesendanger}},
  \bibinfo{journal}{{Nature Comm.}} \textbf{\bibinfo{volume}{{3}}},
  \bibinfo{pages}{{953}} (\bibinfo{year}{{2012}}).

\bibitem[{\citenamefont{Djeghloul et~al.}({2013})\citenamefont{Djeghloul,
  Ibrahim, Cantoni, Bowen, Joly, Boukari, Ohresser, Bertran, Le~Fevre, Thakur
  et~al.}}]{Djeghloul2013}
\bibinfo{author}{\bibfnamefont{F.}~\bibnamefont{Djeghloul}},
  \bibinfo{author}{\bibfnamefont{F.}~\bibnamefont{Ibrahim}},
  \bibinfo{author}{\bibfnamefont{M.}~\bibnamefont{Cantoni}},
  \bibinfo{author}{\bibfnamefont{M.}~\bibnamefont{Bowen}},
  \bibinfo{author}{\bibfnamefont{L.}~\bibnamefont{Joly}},
  \bibinfo{author}{\bibfnamefont{S.}~\bibnamefont{Boukari}},
  \bibinfo{author}{\bibfnamefont{P.}~\bibnamefont{Ohresser}},
  \bibinfo{author}{\bibfnamefont{F.}~\bibnamefont{Bertran}},
  \bibinfo{author}{\bibfnamefont{P.}~\bibnamefont{Le~Fevre}},
  \bibinfo{author}{\bibfnamefont{P.}~\bibnamefont{Thakur}},
  \bibnamefont{et~al.}, \bibinfo{journal}{{Scientific Reports}}
  \textbf{\bibinfo{volume}{{3}}}, \bibinfo{pages}{{1272}}
  (\bibinfo{year}{{2013}}).

\bibitem[{\citenamefont{Atodiresei et~al.}(2010)\citenamefont{Atodiresei,
  Brede, Lazi\ifmmode~\acute{c}\else \'{c}\fi{}, Caciuc, Hoffmann,
  Wiesendanger, and Bl\"ugel}}]{PhysRevLett.105.066601}
\bibinfo{author}{\bibfnamefont{N.}~\bibnamefont{Atodiresei}},
  \bibinfo{author}{\bibfnamefont{J.}~\bibnamefont{Brede}},
  \bibinfo{author}{\bibfnamefont{P.}~\bibnamefont{Lazi\ifmmode~\acute{c}\else
  \'{c}\fi{}}}, \bibinfo{author}{\bibfnamefont{V.}~\bibnamefont{Caciuc}},
  \bibinfo{author}{\bibfnamefont{G.}~\bibnamefont{Hoffmann}},
  \bibinfo{author}{\bibfnamefont{R.}~\bibnamefont{Wiesendanger}},
  \bibnamefont{and} \bibinfo{author}{\bibfnamefont{S.}~\bibnamefont{Bl\"ugel}},
  \bibinfo{journal}{Phys. Rev. Lett.} \textbf{\bibinfo{volume}{105}},
  \bibinfo{pages}{066601} (\bibinfo{year}{2010}).

\bibitem[{\citenamefont{Brede et~al.}(2010)\citenamefont{Brede, Atodiresei,
  Kuck, Lazi\ifmmode~\acute{c}\else \'{c}\fi{}, Caciuc, Morikawa, Hoffmann,
  Bl\"ugel, and Wiesendanger}}]{PhysRevLett.105.047204}
\bibinfo{author}{\bibfnamefont{J.}~\bibnamefont{Brede}},
  \bibinfo{author}{\bibfnamefont{N.}~\bibnamefont{Atodiresei}},
  \bibinfo{author}{\bibfnamefont{S.}~\bibnamefont{Kuck}},
  \bibinfo{author}{\bibfnamefont{P.}~\bibnamefont{Lazi\ifmmode~\acute{c}\else
  \'{c}\fi{}}}, \bibinfo{author}{\bibfnamefont{V.}~\bibnamefont{Caciuc}},
  \bibinfo{author}{\bibfnamefont{Y.}~\bibnamefont{Morikawa}},
  \bibinfo{author}{\bibfnamefont{G.}~\bibnamefont{Hoffmann}},
  \bibinfo{author}{\bibfnamefont{S.}~\bibnamefont{Bl\"ugel}}, \bibnamefont{and}
  \bibinfo{author}{\bibfnamefont{R.}~\bibnamefont{Wiesendanger}},
  \bibinfo{journal}{Phys. Rev. Lett.} \textbf{\bibinfo{volume}{105}},
  \bibinfo{pages}{047204} (\bibinfo{year}{2010}).

\bibitem[{\citenamefont{Zhou et~al.}(2010)\citenamefont{Zhou, Meier, Wiebe, and
  Wiesendanger}}]{PhysRevB.82.012409}
\bibinfo{author}{\bibfnamefont{L.}~\bibnamefont{Zhou}},
  \bibinfo{author}{\bibfnamefont{F.}~\bibnamefont{Meier}},
  \bibinfo{author}{\bibfnamefont{J.}~\bibnamefont{Wiebe}}, \bibnamefont{and}
  \bibinfo{author}{\bibfnamefont{R.}~\bibnamefont{Wiesendanger}},
  \bibinfo{journal}{Phys. Rev. B} \textbf{\bibinfo{volume}{82}},
  \bibinfo{pages}{012409} (\bibinfo{year}{2010}).

\bibitem[{\citenamefont{Ferriani et~al.}(2010)\citenamefont{Ferriani, Lazo, and
  Heinze}}]{Ferriani2010}
\bibinfo{author}{\bibfnamefont{P.}~\bibnamefont{Ferriani}},
  \bibinfo{author}{\bibfnamefont{C.}~\bibnamefont{Lazo}}, \bibnamefont{and}
  \bibinfo{author}{\bibfnamefont{S.}~\bibnamefont{Heinze}},
  \bibinfo{journal}{Phys. Rev. B} \textbf{\bibinfo{volume}{82}},
  \bibinfo{pages}{054411} (\bibinfo{year}{2010}).

\bibitem[{\citenamefont{Wende et~al.}(2007)\citenamefont{Wende, Bernien, Luo,
  Sorg, Ponpandian, Kurde, Miguel, Piantek, Xu, Eckhold et~al.}}]{Wende2007}
\bibinfo{author}{\bibfnamefont{H.}~\bibnamefont{Wende}},
  \bibinfo{author}{\bibfnamefont{M.}~\bibnamefont{Bernien}},
  \bibinfo{author}{\bibfnamefont{J.}~\bibnamefont{Luo}},
  \bibinfo{author}{\bibfnamefont{C.}~\bibnamefont{Sorg}},
  \bibinfo{author}{\bibfnamefont{N.}~\bibnamefont{Ponpandian}},
  \bibinfo{author}{\bibfnamefont{J.}~\bibnamefont{Kurde}},
  \bibinfo{author}{\bibfnamefont{J.}~\bibnamefont{Miguel}},
  \bibinfo{author}{\bibfnamefont{M.}~\bibnamefont{Piantek}},
  \bibinfo{author}{\bibfnamefont{X.}~\bibnamefont{Xu}},
  \bibinfo{author}{\bibfnamefont{P.}~\bibnamefont{Eckhold}},
  \bibnamefont{et~al.}, \bibinfo{journal}{Nature materials}
  \textbf{\bibinfo{volume}{6}}, \bibinfo{pages}{516} (\bibinfo{year}{2007}).

\bibitem[{\citenamefont{Javaid et~al.}(2010)\citenamefont{Javaid, Bowen,
  Boukari, Joly, Beaufrand, Chen, Dappe, Scheurer, Kappler, Arabski
  et~al.}}]{Javaid2010}
\bibinfo{author}{\bibfnamefont{S.}~\bibnamefont{Javaid}},
  \bibinfo{author}{\bibfnamefont{M.}~\bibnamefont{Bowen}},
  \bibinfo{author}{\bibfnamefont{S.}~\bibnamefont{Boukari}},
  \bibinfo{author}{\bibfnamefont{L.}~\bibnamefont{Joly}},
  \bibinfo{author}{\bibfnamefont{J.-B.} \bibnamefont{Beaufrand}},
  \bibinfo{author}{\bibfnamefont{X.}~\bibnamefont{Chen}},
  \bibinfo{author}{\bibfnamefont{Y.~J.} \bibnamefont{Dappe}},
  \bibinfo{author}{\bibfnamefont{F.}~\bibnamefont{Scheurer}},
  \bibinfo{author}{\bibfnamefont{J.-P.} \bibnamefont{Kappler}},
  \bibinfo{author}{\bibfnamefont{J.}~\bibnamefont{Arabski}},
  \bibnamefont{et~al.}, \bibinfo{journal}{Phys. Rev. Lett.}
  \textbf{\bibinfo{volume}{105}}, \bibinfo{pages}{077201}
  (\bibinfo{year}{2010}).

\bibitem[{\citenamefont{Chen and Alouani}(2010)}]{PhysRevB.82.094443}
\bibinfo{author}{\bibfnamefont{X.}~\bibnamefont{Chen}} \bibnamefont{and}
  \bibinfo{author}{\bibfnamefont{M.}~\bibnamefont{Alouani}},
  \bibinfo{journal}{Phys. Rev. B} \textbf{\bibinfo{volume}{82}},
  \bibinfo{pages}{094443} (\bibinfo{year}{2010}).

\bibitem[{\citenamefont{Lach et~al.}(2012)\citenamefont{Lach, Altenhof,
  Tarafder, Schmitt, Ali, Vogel, Sauther, Oppeneer, and Ziegler}}]{Lach2012}
\bibinfo{author}{\bibfnamefont{S.}~\bibnamefont{Lach}},
  \bibinfo{author}{\bibfnamefont{A.}~\bibnamefont{Altenhof}},
  \bibinfo{author}{\bibfnamefont{K.}~\bibnamefont{Tarafder}},
  \bibinfo{author}{\bibfnamefont{F.}~\bibnamefont{Schmitt}},
  \bibinfo{author}{\bibfnamefont{M.~E.} \bibnamefont{Ali}},
  \bibinfo{author}{\bibfnamefont{M.}~\bibnamefont{Vogel}},
  \bibinfo{author}{\bibfnamefont{J.}~\bibnamefont{Sauther}},
  \bibinfo{author}{\bibfnamefont{P.~M.} \bibnamefont{Oppeneer}},
  \bibnamefont{and} \bibinfo{author}{\bibfnamefont{C.}~\bibnamefont{Ziegler}},
  \bibinfo{journal}{Advanced Functional Materials}
  \textbf{\bibinfo{volume}{22}}, \bibinfo{pages}{989} (\bibinfo{year}{2012}).

\bibitem[{\citenamefont{Raman et~al.}(2013)\citenamefont{Raman, Kamerbeek,
  Mukherjee, Atodiresei, Sen, Lazi\'{c}, Caciuc, Michel, Stalke, Mandal
  et~al.}}]{Raman2013}
\bibinfo{author}{\bibfnamefont{K.~V.} \bibnamefont{Raman}},
  \bibinfo{author}{\bibfnamefont{A.~M.} \bibnamefont{Kamerbeek}},
  \bibinfo{author}{\bibfnamefont{A.}~\bibnamefont{Mukherjee}},
  \bibinfo{author}{\bibfnamefont{N.}~\bibnamefont{Atodiresei}},
  \bibinfo{author}{\bibfnamefont{T.~K.} \bibnamefont{Sen}},
  \bibinfo{author}{\bibfnamefont{P.}~\bibnamefont{Lazi\'{c}}},
  \bibinfo{author}{\bibfnamefont{V.}~\bibnamefont{Caciuc}},
  \bibinfo{author}{\bibfnamefont{R.}~\bibnamefont{Michel}},
  \bibinfo{author}{\bibfnamefont{D.}~\bibnamefont{Stalke}},
  \bibinfo{author}{\bibfnamefont{S.~K.} \bibnamefont{Mandal}},
  \bibnamefont{et~al.}, \bibinfo{journal}{Nature}
  \textbf{\bibinfo{volume}{493}}, \bibinfo{pages}{509} (\bibinfo{year}{2013}).

\bibitem[{\citenamefont{Annese et~al.}(2013)\citenamefont{Annese, Casolari,
  Fujii, and Rossi}}]{PhysRevB.87.054420}
\bibinfo{author}{\bibfnamefont{E.}~\bibnamefont{Annese}},
  \bibinfo{author}{\bibfnamefont{F.}~\bibnamefont{Casolari}},
  \bibinfo{author}{\bibfnamefont{J.}~\bibnamefont{Fujii}}, \bibnamefont{and}
  \bibinfo{author}{\bibfnamefont{G.}~\bibnamefont{Rossi}},
  \bibinfo{journal}{Phys. Rev. B} \textbf{\bibinfo{volume}{87}},
  \bibinfo{pages}{054420} (\bibinfo{year}{2013}).

\bibitem[{\citenamefont{Lodi~Rizzini et~al.}({2012})\citenamefont{Lodi~Rizzini,
  Krull, Balashov, Mugarza, Nistor, Yakhou, Sessi, Klyatskaya, Ruben, Stepanow
  et~al.}}]{Rizzini2012}
\bibinfo{author}{\bibfnamefont{A.}~\bibnamefont{Lodi~Rizzini}},
  \bibinfo{author}{\bibfnamefont{C.}~\bibnamefont{Krull}},
  \bibinfo{author}{\bibfnamefont{T.}~\bibnamefont{Balashov}},
  \bibinfo{author}{\bibfnamefont{A.}~\bibnamefont{Mugarza}},
  \bibinfo{author}{\bibfnamefont{C.}~\bibnamefont{Nistor}},
  \bibinfo{author}{\bibfnamefont{F.}~\bibnamefont{Yakhou}},
  \bibinfo{author}{\bibfnamefont{V.}~\bibnamefont{Sessi}},
  \bibinfo{author}{\bibfnamefont{S.}~\bibnamefont{Klyatskaya}},
  \bibinfo{author}{\bibfnamefont{M.}~\bibnamefont{Ruben}},
  \bibinfo{author}{\bibfnamefont{S.}~\bibnamefont{Stepanow}},
  \bibnamefont{et~al.}, \bibinfo{journal}{{Nano Letters}}
  \textbf{\bibinfo{volume}{{12}}}, \bibinfo{pages}{{5703}}
  (\bibinfo{year}{{2012}}).

\bibitem[{\citenamefont{Bagrets et~al.}({2012})\citenamefont{Bagrets, Schmaus,
  Jaafar, Kramczynski, Yamada, Alouani, Wulfhekel, and Evers}}]{Bagrets2012}
\bibinfo{author}{\bibfnamefont{A.}~\bibnamefont{Bagrets}},
  \bibinfo{author}{\bibfnamefont{S.}~\bibnamefont{Schmaus}},
  \bibinfo{author}{\bibfnamefont{A.}~\bibnamefont{Jaafar}},
  \bibinfo{author}{\bibfnamefont{D.}~\bibnamefont{Kramczynski}},
  \bibinfo{author}{\bibfnamefont{T.~K.} \bibnamefont{Yamada}},
  \bibinfo{author}{\bibfnamefont{M.}~\bibnamefont{Alouani}},
  \bibinfo{author}{\bibfnamefont{W.}~\bibnamefont{Wulfhekel}},
  \bibnamefont{and} \bibinfo{author}{\bibfnamefont{F.}~\bibnamefont{Evers}},
  \bibinfo{journal}{{Nano Letters}} \textbf{\bibinfo{volume}{{12}}},
  \bibinfo{pages}{{5131}} (\bibinfo{year}{{2012}}).

\bibitem[{\citenamefont{Bode et~al.}(2002)\citenamefont{Bode, Heinze, Kubetzka,
  Pietzsch, Hennefarth, Getzlaff, Wiesendanger, Nie, Bihlmayer, and
  Bl\"ugel}}]{PhysRevB.66.014425}
\bibinfo{author}{\bibfnamefont{M.}~\bibnamefont{Bode}},
  \bibinfo{author}{\bibfnamefont{S.}~\bibnamefont{Heinze}},
  \bibinfo{author}{\bibfnamefont{A.}~\bibnamefont{Kubetzka}},
  \bibinfo{author}{\bibfnamefont{O.}~\bibnamefont{Pietzsch}},
  \bibinfo{author}{\bibfnamefont{M.}~\bibnamefont{Hennefarth}},
  \bibinfo{author}{\bibfnamefont{M.}~\bibnamefont{Getzlaff}},
  \bibinfo{author}{\bibfnamefont{R.}~\bibnamefont{Wiesendanger}},
  \bibinfo{author}{\bibfnamefont{X.}~\bibnamefont{Nie}},
  \bibinfo{author}{\bibfnamefont{G.}~\bibnamefont{Bihlmayer}},
  \bibnamefont{and} \bibinfo{author}{\bibfnamefont{S.}~\bibnamefont{Bl\"ugel}},
  \bibinfo{journal}{Phys. Rev. B} \textbf{\bibinfo{volume}{66}},
  \bibinfo{pages}{014425} (\bibinfo{year}{2002}).

\bibitem[{\citenamefont{Bode et~al.}({2007})\citenamefont{Bode, Heide, von
  Bergmann, Ferriani, Heinze, Bihlmayer, Kubetzka, Pietzsch, Bl\"ugel, and
  Wiesendanger}}]{Bode2007}
\bibinfo{author}{\bibfnamefont{M.}~\bibnamefont{Bode}},
  \bibinfo{author}{\bibfnamefont{M.}~\bibnamefont{Heide}},
  \bibinfo{author}{\bibfnamefont{K.}~\bibnamefont{von Bergmann}},
  \bibinfo{author}{\bibfnamefont{P.}~\bibnamefont{Ferriani}},
  \bibinfo{author}{\bibfnamefont{S.}~\bibnamefont{Heinze}},
  \bibinfo{author}{\bibfnamefont{G.}~\bibnamefont{Bihlmayer}},
  \bibinfo{author}{\bibfnamefont{A.}~\bibnamefont{Kubetzka}},
  \bibinfo{author}{\bibfnamefont{O.}~\bibnamefont{Pietzsch}},
  \bibinfo{author}{\bibfnamefont{S.}~\bibnamefont{Bl\"ugel}}, \bibnamefont{and}
  \bibinfo{author}{\bibfnamefont{R.}~\bibnamefont{Wiesendanger}},
  \bibinfo{journal}{{Nature}} \textbf{\bibinfo{volume}{{447}}},
  \bibinfo{pages}{{190}} (\bibinfo{year}{{2007}}).

\bibitem[{\citenamefont{Serrate et~al.}({2010})\citenamefont{Serrate, Ferriani,
  Yoshida, Hla, Menzel, von Bergmann, Heinze, Kubetzka, and
  Wiesendanger}}]{Serrate2010}
\bibinfo{author}{\bibfnamefont{D.}~\bibnamefont{Serrate}},
  \bibinfo{author}{\bibfnamefont{P.}~\bibnamefont{Ferriani}},
  \bibinfo{author}{\bibfnamefont{Y.}~\bibnamefont{Yoshida}},
  \bibinfo{author}{\bibfnamefont{S.-W.} \bibnamefont{Hla}},
  \bibinfo{author}{\bibfnamefont{M.}~\bibnamefont{Menzel}},
  \bibinfo{author}{\bibfnamefont{K.}~\bibnamefont{von Bergmann}},
  \bibinfo{author}{\bibfnamefont{S.}~\bibnamefont{Heinze}},
  \bibinfo{author}{\bibfnamefont{A.}~\bibnamefont{Kubetzka}}, \bibnamefont{and}
  \bibinfo{author}{\bibfnamefont{R.}~\bibnamefont{Wiesendanger}},
  \bibinfo{journal}{{Nature Nanotechnology}} \textbf{\bibinfo{volume}{{5}}},
  \bibinfo{pages}{{350}} (\bibinfo{year}{{2010}}).

\bibitem[{\citenamefont{Wortmann et~al.}(2001)\citenamefont{Wortmann, Heinze,
  Kurz, Bihlmayer, and Bl\"ugel}}]{PhysRevLett.86.4132}
\bibinfo{author}{\bibfnamefont{D.}~\bibnamefont{Wortmann}},
  \bibinfo{author}{\bibfnamefont{S.}~\bibnamefont{Heinze}},
  \bibinfo{author}{\bibfnamefont{P.}~\bibnamefont{Kurz}},
  \bibinfo{author}{\bibfnamefont{G.}~\bibnamefont{Bihlmayer}},
  \bibnamefont{and} \bibinfo{author}{\bibfnamefont{S.}~\bibnamefont{Bl\"ugel}},
  \bibinfo{journal}{Phys. Rev. Lett.} \textbf{\bibinfo{volume}{86}},
  \bibinfo{pages}{4132} (\bibinfo{year}{2001}).

\bibitem[{\citenamefont{Kresse and Furthm\"uller}(1996)}]{Kresse1996}
\bibinfo{author}{\bibfnamefont{G.}~\bibnamefont{Kresse}} \bibnamefont{and}
  \bibinfo{author}{\bibfnamefont{J.}~\bibnamefont{Furthm\"uller}},
  \bibinfo{journal}{Phys. Rev. B} \textbf{\bibinfo{volume}{54}},
  \bibinfo{pages}{11169} (\bibinfo{year}{1996}).

\bibitem[{\citenamefont{Kresse and Joubert}(1999)}]{Kresse1999}
\bibinfo{author}{\bibfnamefont{G.}~\bibnamefont{Kresse}} \bibnamefont{and}
  \bibinfo{author}{\bibfnamefont{D.}~\bibnamefont{Joubert}},
  \bibinfo{journal}{Phys. Rev. B} \textbf{\bibinfo{volume}{59}},
  \bibinfo{pages}{1758} (\bibinfo{year}{1999}).

\bibitem[{\citenamefont{Perdew et~al.}(1996)\citenamefont{Perdew, Burke, and
  Ernzerhof}}]{Perdew1996}
\bibinfo{author}{\bibfnamefont{J.~P.} \bibnamefont{Perdew}},
  \bibinfo{author}{\bibfnamefont{K.}~\bibnamefont{Burke}}, \bibnamefont{and}
  \bibinfo{author}{\bibfnamefont{M.}~\bibnamefont{Ernzerhof}},
  \bibinfo{journal}{Phys. Rev. Lett.} \textbf{\bibinfo{volume}{77}},
  \bibinfo{pages}{3865} (\bibinfo{year}{1996}).

\bibitem[{\citenamefont{Bl\"ochl}(1994)}]{PhysRevB.50.17953}
\bibinfo{author}{\bibfnamefont{P.~E.} \bibnamefont{Bl\"ochl}},
  \bibinfo{journal}{Phys. Rev. B} \textbf{\bibinfo{volume}{50}},
  \bibinfo{pages}{17953} (\bibinfo{year}{1994}).

\bibitem[{\citenamefont{Monkhorst and Pack}(1976)}]{PhysRevB.13.5188}
\bibinfo{author}{\bibfnamefont{H.~J.} \bibnamefont{Monkhorst}}
  \bibnamefont{and} \bibinfo{author}{\bibfnamefont{J.~D.} \bibnamefont{Pack}},
  \bibinfo{journal}{Phys. Rev. B} \textbf{\bibinfo{volume}{13}},
  \bibinfo{pages}{5188} (\bibinfo{year}{1976}).

\bibitem[{\citenamefont{Tersoff and Hamann}(1985)}]{PhysRevB.31.805}
\bibinfo{author}{\bibfnamefont{J.}~\bibnamefont{Tersoff}} \bibnamefont{and}
  \bibinfo{author}{\bibfnamefont{D.~R.} \bibnamefont{Hamann}},
  \bibinfo{journal}{Phys. Rev. B} \textbf{\bibinfo{volume}{31}},
  \bibinfo{pages}{805} (\bibinfo{year}{1985}).

\bibitem[{\citenamefont{Grimme}(2006)}]{Grimme2006}
\bibinfo{author}{\bibfnamefont{S.}~\bibnamefont{Grimme}},
  \bibinfo{journal}{Journal of Computational Chemistry}
  \textbf{\bibinfo{volume}{27}}, \bibinfo{pages}{1787} (\bibinfo{year}{2006}).

\bibitem[{\citenamefont{Bučko et~al.}(2010)\citenamefont{Bučko, Hafner,
  Lebègue, and Ángyán}}]{Bucko2010}
\bibinfo{author}{\bibfnamefont{T.}~\bibnamefont{Bučko}},
  \bibinfo{author}{\bibfnamefont{J.}~\bibnamefont{Hafner}},
  \bibinfo{author}{\bibfnamefont{S.}~\bibnamefont{Lebègue}},
  \bibnamefont{and} \bibinfo{author}{\bibfnamefont{J.~G.}
  \bibnamefont{Ángyán}}, \bibinfo{journal}{The Journal of Physical
  Chemistry A} \textbf{\bibinfo{volume}{114}}, \bibinfo{pages}{11814}
  (\bibinfo{year}{2010}).

\bibitem[{\citenamefont{Kaufman et~al.}({1948})\citenamefont{Kaufman,
  Fankuchen, and Mark}}]{Kaufman1948}
\bibinfo{author}{\bibfnamefont{H.}~\bibnamefont{Kaufman}},
  \bibinfo{author}{\bibfnamefont{I.}~\bibnamefont{Fankuchen}},
  \bibnamefont{and} \bibinfo{author}{\bibfnamefont{H.}~\bibnamefont{Mark}},
  \bibinfo{journal}{{Nature}} \textbf{\bibinfo{volume}{{161}}},
  \bibinfo{pages}{{165}} (\bibinfo{year}{{1948}}).

\bibitem[{\citenamefont{Harutyunyan et~al.}(2013)\citenamefont{Harutyunyan,
  Callsen, Allmers, Caciuc, Bl\"{u}gel, Atodiresei, and
  Wegner}}]{Harutyunyan2013}
\bibinfo{author}{\bibfnamefont{H.}~\bibnamefont{Harutyunyan}},
  \bibinfo{author}{\bibfnamefont{M.}~\bibnamefont{Callsen}},
  \bibinfo{author}{\bibfnamefont{T.}~\bibnamefont{Allmers}},
  \bibinfo{author}{\bibfnamefont{V.}~\bibnamefont{Caciuc}},
  \bibinfo{author}{\bibfnamefont{S.}~\bibnamefont{Bl\"{u}gel}},
  \bibinfo{author}{\bibfnamefont{N.}~\bibnamefont{Atodiresei}},
  \bibnamefont{and} \bibinfo{author}{\bibfnamefont{D.}~\bibnamefont{Wegner}},
  \bibinfo{journal}{Chemical Communications} \textbf{\bibinfo{volume}{49}},
  \bibinfo{pages}{5993} (\bibinfo{year}{2013}).

\bibitem[{\citenamefont{Jaeger et~al.}({2004})\citenamefont{Jaeger, van
  Heijnsbergen, Klippenstein, von Helden, Meijer, and Duncan}}]{Jaeger2004}
\bibinfo{author}{\bibfnamefont{T.}~\bibnamefont{Jaeger}},
  \bibinfo{author}{\bibfnamefont{D.}~\bibnamefont{van Heijnsbergen}},
  \bibinfo{author}{\bibfnamefont{S.}~\bibnamefont{Klippenstein}},
  \bibinfo{author}{\bibfnamefont{G.}~\bibnamefont{von Helden}},
  \bibinfo{author}{\bibfnamefont{G.}~\bibnamefont{Meijer}}, \bibnamefont{and}
  \bibinfo{author}{\bibfnamefont{M.}~\bibnamefont{Duncan}},
  \bibinfo{journal}{{Journal of the American Chemical Society}}
  \textbf{\bibinfo{volume}{{126}}}, \bibinfo{pages}{{10981}}
  (\bibinfo{year}{{2004}}).

\bibitem[{\citenamefont{Duncan}({2008})}]{Duncan2008}
\bibinfo{author}{\bibfnamefont{M.~A.} \bibnamefont{Duncan}},
  \bibinfo{journal}{{International Journal of Mass Spectrometry}}
  \textbf{\bibinfo{volume}{{272}}}, \bibinfo{pages}{{99}}
  (\bibinfo{year}{{2008}}).

\bibitem[{\citenamefont{Kandalam et~al.}(2004)\citenamefont{Kandalam, Rao,
  Jena, and Pandey}}]{kandalam:10414}
\bibinfo{author}{\bibfnamefont{A.~K.} \bibnamefont{Kandalam}},
  \bibinfo{author}{\bibfnamefont{B.~K.} \bibnamefont{Rao}},
  \bibinfo{author}{\bibfnamefont{P.}~\bibnamefont{Jena}}, \bibnamefont{and}
  \bibinfo{author}{\bibfnamefont{R.}~\bibnamefont{Pandey}},
  \bibinfo{journal}{The Journal of Chemical Physics}
  \textbf{\bibinfo{volume}{120}}, \bibinfo{pages}{10414}
  (\bibinfo{year}{2004}).

\bibitem[{\citenamefont{Mokrousov et~al.}({2007})\citenamefont{Mokrousov,
  Atodiresei, Bihlmayer, Heinze, and Bl\"ugel}}]{Mokrousov2007}
\bibinfo{author}{\bibfnamefont{Y.}~\bibnamefont{Mokrousov}},
  \bibinfo{author}{\bibfnamefont{N.}~\bibnamefont{Atodiresei}},
  \bibinfo{author}{\bibfnamefont{G.}~\bibnamefont{Bihlmayer}},
  \bibinfo{author}{\bibfnamefont{S.}~\bibnamefont{Heinze}}, \bibnamefont{and}
  \bibinfo{author}{\bibfnamefont{S.}~\bibnamefont{Bl\"ugel}},
  \bibinfo{journal}{{Nanotechnology}} \textbf{\bibinfo{volume}{{18}}},
  \bibinfo{pages}{{495402}} (\bibinfo{year}{{2007}}).

\bibitem[{\citenamefont{Henkelman et~al.}({2006})\citenamefont{Henkelman,
  Arnaldsson, and Jonsson}}]{Henkelman2006}
\bibinfo{author}{\bibfnamefont{G.}~\bibnamefont{Henkelman}},
  \bibinfo{author}{\bibfnamefont{A.}~\bibnamefont{Arnaldsson}},
  \bibnamefont{and} \bibinfo{author}{\bibfnamefont{H.}~\bibnamefont{Jonsson}},
  \bibinfo{journal}{{Computational Materials Science}}
  \textbf{\bibinfo{volume}{{36}}}, \bibinfo{pages}{{354}}
  (\bibinfo{year}{{2006}}).

\bibitem[{\citenamefont{Atodiresei et~al.}(2011)\citenamefont{Atodiresei,
  Caciuc, Lazi\ifmmode~\acute{c}\else \'{c}\fi{}, and
  Bl\"ugel}}]{PhysRevB.84.172402}
\bibinfo{author}{\bibfnamefont{N.}~\bibnamefont{Atodiresei}},
  \bibinfo{author}{\bibfnamefont{V.}~\bibnamefont{Caciuc}},
  \bibinfo{author}{\bibfnamefont{P.}~\bibnamefont{Lazi\ifmmode~\acute{c}\else
  \'{c}\fi{}}}, \bibnamefont{and}
  \bibinfo{author}{\bibfnamefont{S.}~\bibnamefont{Bl\"ugel}},
  \bibinfo{journal}{Phys. Rev. B} \textbf{\bibinfo{volume}{84}},
  \bibinfo{pages}{172402} (\bibinfo{year}{2011}).

\bibitem[{\citenamefont{Methfessel et~al.}(2011)\citenamefont{Methfessel,
  Steil, Baadji, Gro\ss{}mann, Koffler, Sanvito, Aeschlimann, Cinchetti, and
  Elmers}}]{PhysRevB.84.224403}
\bibinfo{author}{\bibfnamefont{T.}~\bibnamefont{Methfessel}},
  \bibinfo{author}{\bibfnamefont{S.}~\bibnamefont{Steil}},
  \bibinfo{author}{\bibfnamefont{N.}~\bibnamefont{Baadji}},
  \bibinfo{author}{\bibfnamefont{N.}~\bibnamefont{Gro\ss{}mann}},
  \bibinfo{author}{\bibfnamefont{K.}~\bibnamefont{Koffler}},
  \bibinfo{author}{\bibfnamefont{S.}~\bibnamefont{Sanvito}},
  \bibinfo{author}{\bibfnamefont{M.}~\bibnamefont{Aeschlimann}},
  \bibinfo{author}{\bibfnamefont{M.}~\bibnamefont{Cinchetti}},
  \bibnamefont{and} \bibinfo{author}{\bibfnamefont{H.~J.}
  \bibnamefont{Elmers}}, \bibinfo{journal}{Phys. Rev. B}
  \textbf{\bibinfo{volume}{84}}, \bibinfo{pages}{224403}
  (\bibinfo{year}{2011}).

\bibitem[{\citenamefont{Wang et~al.}(2013)\citenamefont{Wang, Zhu, Manchon, and
  Schwingenschlogl}}]{Wang2013}
\bibinfo{author}{\bibfnamefont{X.}~\bibnamefont{Wang}},
  \bibinfo{author}{\bibfnamefont{Z.}~\bibnamefont{Zhu}},
  \bibinfo{author}{\bibfnamefont{A.}~\bibnamefont{Manchon}}, \bibnamefont{and}
  \bibinfo{author}{\bibfnamefont{U.}~\bibnamefont{Schwingenschlogl}},
  \bibinfo{journal}{Applied Physics Letters} \textbf{\bibinfo{volume}{102}},
  \bibinfo{pages}{111604} (\bibinfo{year}{2013}).

\bibitem[{\citenamefont{Momma and Izumi}({2011})}]{vesta}
\bibinfo{author}{\bibfnamefont{K.}~\bibnamefont{Momma}} \bibnamefont{and}
  \bibinfo{author}{\bibfnamefont{F.}~\bibnamefont{Izumi}},
  \bibinfo{journal}{{Journal of Applied Crystallography}}
  \textbf{\bibinfo{volume}{{44}}}, \bibinfo{pages}{{1272}}
  (\bibinfo{year}{{2011}}).

\end{thebibliography}
\end{document}